\documentclass[a4paper,aps,prd,10pt,preprintnumbers,showpacs,twocolumn,superscriptaddress,nofootinbib,amsmath,amssymb]{revtex4-1}
\usepackage{graphicx,color}
\usepackage{cmap}
\usepackage[utf8]{inputenc}
\usepackage[T1]{fontenc}

\def\re#1{Re(#1)}
\def\im#1{Im(#1)}

\def\Order#1{{\cal O}\left(#1\right)}

\begin{document}
\title{Bardeen spacetime as quantum corrected black hole: Grey-body factors and quasinormal modes of gravitational perturbations}
\author{Bekir Can L{\"u}tf{\"u}o{\u g}lu}
\email{bekir.lutfuoglu@uhk.cz}
\affiliation{Department of Physics, Faculty of Science, University of Hradec Králové, Rokitanského 62/26, 500 03 Hradec Králové, Czech Republic}

\author{Javlon~Rayimbaev} 
\email{javlon@astrin.uz}
\affiliation{Institute of Theoretical Physics, National University of Uzbekistan, Tashkent 100174, Uzbekistan}
\affiliation{University of Tashkent for Applied Sciences, Gavhar Str. 1, Tashkent 700127, Uzbekistan}

\author{Sardor~Murodov} 
\email{mursardor@ifar.uz}
\affiliation{New Uzbekistan University, Movarounnahr Str. 1, Tashkent 100000, Uzbekistan}
\affiliation{Tashkent State Technical University, Tashkent 100095,Uzbekistan}

\author{Jakhongir~Kurbanov} \email{jaxongir0903@gmail.com} \affiliation{Kimyo International University in Tashkent, Shota Rustaveli Street 156, Tashkent 100121, Uzbekistan}

\author{Muhammad~Matyoqubov} 
\email{m_matyoqubov@mamunedu.uz} \affiliation{Mamun University, Bolkhovuz Street 2, Khiva 220900, Uzbekistan}
\begin{abstract}
We study axial gravitational perturbations of the asymptotically flat Bardeen spacetime interpreted as a string-T-duality-inspired quantum-corrected Schwarzschild black hole. Starting from the anisotropic-fluid background, we derive the Regge--Wheeler-type master equation and the corresponding effective potential, and compute quasinormal modes with high-order WKB--Pad\'e and time-domain methods. We show that increasing the quantum-correction scale $\ell_0$ raises and inwardly shifts the barrier, causing the black hole to ring at higher frequencies and decay more slowly. The same deformation suppresses low-frequency transmission, shifts the onset of grey-body factors to larger frequencies, and reorganizes the partial and total absorption cross-sections. Overall, the results identify a clear and consistent imprint of short-distance regularization on both ringdown and scattering observables.
\end{abstract}
\maketitle

\section{Introduction}

The ringdown stage of a perturbed black hole is governed by a discrete set of
complex frequencies known as quasinormal modes (QNMs), whose real and
imaginary parts determine the oscillation frequency and damping time,
respectively. Because these modes are fixed by the background geometry and by
the spin of the perturbing field, they provide one of the cleanest probes of
the near-horizon region and of possible deviations from the classical black
hole paradigm. In the era of gravitational-wave astronomy, QNMs have therefore
become central both to black-hole spectroscopy and to precision tests of strong
gravity, since the observed ringdown signal is directly sensitive to the
underlying effective potential that governs perturbations
\cite{Abbott2016,Kokkotas1999,Berti2009,Bolokhov:2025uxz}. This makes the gravitational QNM
spectrum especially valuable for quantum-corrected geometries, where even a
small deformation of the potential barrier may leave the fundamental mode only
slightly changed while substantially affecting overtones and damping rates \cite{Konoplya:2022pbc}.

Grey-body factors (GBFs) provide complementary information. Hawking radiation
is created near the event horizon, but the flux measured by an asymptotic
observer is filtered by the curvature-induced potential barrier surrounding the
black hole \cite{Kanti:2002nr,Page:1976df,Page:1976ki}. As a result, the emitted spectrum is not exactly Planckian; rather,
it is weighted by transmission coefficients that depend on frequency, angular
momentum, and field spin \cite{Hawking1975}. GBFs therefore govern
absorption probabilities, scattering amplitudes, and Hawking energy-emission
rates, and they furnish a natural bridge between black-hole thermodynamics and
wave propagation. Since QNMs and GBFs are determined by the same effective
potential, albeit with different boundary conditions, studying both quantities
together yields a considerably more complete characterization of a black hole
than either observable alone {\color{black}\cite{Konoplya2023,Skvortsova2025,Lutfuoglu2026,Konoplya:2026qql}}.

Regular black holes are particularly attractive laboratories for such studies.
They offer a phenomenological route toward resolving the spacetime singularity
problem while preserving horizons and much of the familiar black-hole exterior.
The Bardeen solution occupies a distinguished place in this program as the
first regular black-hole geometry proposed in the literature \cite{Bardeen1968}.
It was later shown that the Bardeen metric can be supported by nonlinear
electrodynamics, thereby endowing the model with a concrete matter source and a
consistent dynamical interpretation \cite{Ayon1998,Ayon2000}. {\color{black}
Regular geometries have been used to investigate quasinormal spectra and
scattering \cite{Rincon2020,Dey2019,Zhao2024,Bonanno:2025dry,Lin:2013ofa,Flachi:2012nv,Konoplya:2024kih,Fernando:2012yw,Lopez:2022uie,Konoplya:2025uta}.
Perturbations in the nonlinear-electrodynamic interpretation are considered
in Refs.~\cite{Toshmatov:2018ell,Toshmatov:2019gxg}.
Studies of matter environments and alternative regular models provide
further examples \cite{Al-Badawi:2023lke,Pedraza:2021hzw,DuttaRoy:2022ytr,Skvortsova:2026unq,Bolokhov:2025fto,Konoplya:2025ect,Saka:2025xxl},
alongside nonsingular conformal, effective and quantum-motivated geometries
\cite{Toshmatov:2017bpx,Konoplya:2023ppx,Mukohyama:2023xyf,Konoplya:2022hll,Konoplya:2023aph,Gingrich:2024tuf,Zhang:2024nny,Konoplya:2026gim,Dubinsky:2026wcv,Rahmatov:2026ssl,  Lutfuoglu:2026fpx}.
Their connections with optical or thermodynamic observables have also been explored \cite{Held:2019xde,MahdavianYekta:2019pol,Cai:2021ele,Jusufi:2020odz,Jawad:2020hju,Guo:2024jhg,Huang:2023aet,Errehymy:2026fgm, Rahmatov:2026twe}.
Related weak-lensing analyses in modified gravity are presented in
Ref.~\cite{Rahmatov:2026mdo}, while nonsingular anisotropic stellar
configurations are discussed in Ref.~\cite{Asghar:2026que}.
Echoes, long-lived modes and Bardeen-family radiation supply complementary
probes of these geometries \cite{Yang:2021cvh,Skvortsova:2025cah,Konoplya2023,Bolokhov2024,Li:2014fka}.
}

An especially appealing twist was proposed by Nicolini, Spallucci, and
Wondrak, who showed that a neutral, regular black-hole solution obtained from a
string-T-duality-inspired zero-point-length modification of the static
potential is formally equivalent to the Bardeen geometry, up to the precise
interpretation of the ultraviolet regulator \cite{Nicolini2019}. In this
picture, the Bardeen spacetime can be read not merely as a nonlinear
electrodynamics black hole, but also as an effective quantum-corrected
Schwarzschild solution. This reinterpretation is important for phenomenology:
it turns the Bardeen parameter into an economical proxy for short-distance
quantum effects and provides a simple, regular geometry in which one can ask
how quantum corrections modify classical observables without giving up analytic
control over the background.

The perturbative properties of this spacetime have already attracted
considerable attention, but mainly in the test-field sector. In particular,
Konoplya, Ovchinnikov, and Ahmedov revisited the Bardeen spacetime in its
quantum-corrected interpretation and computed accurate scalar,
electromagnetic, and Dirac QNMs, together with GBFs and Hawking
emission rates \cite{Konoplya2023}. Their analysis showed that near-horizon
quantum corrections can produce a pronounced outburst of overtones even when
the fundamental mode stays close to its Schwarzschild counterpart, and that the
same corrections substantially suppress Hawking radiation. Soon afterwards,
Bolokhov investigated massive scalar perturbations of the same background and
found quasiresonances, long-lived modes, and oscillatory late-time tails
\cite{Bolokhov2024}. In parallel, related Bardeen-family geometries were
studied in several settings: gravitational perturbations of the Bardeen black
hole surrounded by quintessence were analyzed in Ref.~\cite{Saleh2018};
electromagnetic and gravitational QNMs together with GBFs were investigated for
the Bardeen--de Sitter spacetime in Ref.~\cite{Dey2019}; and the axial/polar
gravitational sectors and their isospectrality properties were studied for
Bardeen (anti-)de Sitter black holes in Ref.~\cite{Zhao2024}. More broadly,
recent work on quantum-corrected or regular black holes has reinforced the idea
that short-distance modifications may leave robust signatures in higher
overtones, absorption spectra, and the interplay between QNMs and GBFs
\cite{KonoplyaStashko2025,Lutfuoglu2025,Skvortsova2025,Lutfuoglu2026,Shi2026}.

Despite this progress, a dedicated study of gravitational QNMs and
GBFs for the asymptotically flat Bardeen spacetime in the specific
string-T-duality-inspired quantum-corrected interpretation is still missing.
This gap is especially noteworthy because the gravitational sector is the most
directly relevant one for black-hole spectroscopy: it determines the ringdown
signal itself and controls the spin-2 transmission probabilities associated
with graviton scattering and Hawking emission. The purpose of the present work
is to fill this gap by deriving the gravitational perturbation equations for
this background and by analyzing the corresponding QNM spectrum and grey-body
factors. In this way, one can disentangle which signatures arise generically
from regularity and which are genuinely tied to the quantum-corrected
interpretation of the Bardeen geometry.

The remainder of this paper is organized as follows. In Sec.~\ref{sec:metric} we review the quantum-corrected Bardeen geometry and summarize its basic horizon and regularity properties. In Sec.~\ref{sec:axial} we derive the master equation and the corresponding axial effective potential for gravitational perturbations. The WKB and time-domain methods used in the analysis are outlined in Secs.~\ref{sec:wkb} and \ref{sec:time}, respectively. Section~\ref{sec:qnm} presents the gravitational quasinormal spectra and their time-domain interpretation, while Sec.~\ref{sec:gbf} discusses the GBFs together with the partial and total absorption cross-sections (ACSs). Finally, Sec.~\ref{sec:conclusions} summarizes the main results and their physical implications.

\section{Quantum-corrected Bardeen spacetime}\label{sec:metric}

In the interpretation advocated by Nicolini, Spallucci, and Wondrak, the
geometry under consideration is not viewed as a magnetic monopole solution of
nonlinear electrodynamics, but rather as an effective neutral black-hole metric
emerging from string T-duality through its connection with path-integral
self-duality and the associated zero-point length of spacetime
\cite{Padmanabhan1998,Nicolini2019}. In this picture, the short-distance
structure of the propagator is modified in a nonperturbative way, so that the
static Newtonian potential becomes regular instead of diverging at the origin.
When this regularized {procedure is introduced,} one obtains a metric that is formally equivalent
to the Bardeen spacetime \cite{Bardeen1968,Ayon2000}, but whose deformation
parameter is reinterpreted as a quantum-correction scale rather than as a
magnetic charge \cite{Nicolini2019,Konoplya2023}.

{The starting point is the observation that T-duality implies the existence of a fundamental zero-point length, \({\color{black}\ell_0}\), which regularizes the ultraviolet behavior of the propagator. This effect modifies the Newtonian potential to
\begin{equation}\label{eq:regularized_potential}
\Phi(r)=-\frac{M}{\sqrt{r^2+{\color{black}\ell_0}^2}},
\end{equation}
which reduces to the classical Newtonian potential in the limit \({\color{black}\ell_0}\to0\). }

{\color{black}
The regularized Newtonian potential is not identified directly with
$(f-1)/2$. Instead, in geometrized units, the flat-space Poisson equation
for $\Phi(r)=-M/\sqrt{r^2+\ell_0^2}$ defines the effective density,
and its enclosed mass fixes the metric:
\begin{align*}
\rho_{\rm eff}(r)&=\frac{\nabla^2\Phi}{4\pi}
=\frac{3M\ell_0^2}{4\pi(r^2+\ell_0^2)^{5/2}},\\
m(r)&=4\pi\int_0^r\rho_{\rm eff}(s)s^2\,ds
=\frac{Mr^3}{(r^2+\ell_0^2)^{3/2}}.
\end{align*}
Here the regular integration constant is fixed by $m(0)=0$, and
$f=1-2m(r)/r$ gives the Bardeen lapse below, rather than $f=1+2\Phi$.
This is the density and mass construction of Ref.~\cite{Nicolini2019};
the pressures completing the conserved relativistic source are given in
Sec.~\ref{sec:axial}.
}

{\color{black}
The resulting static, spherically symmetric line element is
}
\begin{equation}\label{metric}
 ds^2=-f(r)dt^2+\frac{dr^2}{f(r)}+r^2\left(d\theta^2+\sin^2\theta\,d\phi^2\right),
\end{equation}
with metric function
\begin{equation}\label{eq:bardeen_metric_function}
 f(r)=1-\frac{2 M r^2}{\left(r^2+\ell_0^2\right)^{3/2}},
\end{equation}
where $M$ is the asymptotic mass and $\ell_0$ denotes the zero-point-length
parameter induced by the underlying T-duality scenario. Following
Ref.~\cite{Nicolini2019}, one may relate this parameter to the Regge slope by
$\ell_0=2\pi\sqrt{\alpha'}$, so that $\ell_0$ is expected to be of the order
of the Planck length. Since the geometry depends only on $\ell_0^2$, it is
sufficient to restrict attention to $\ell_0\geq 0$.

The basic properties of the metric are immediate from its asymptotic
expansions. At large radial distance one finds
\begin{equation}
 f(r)=1-\frac{2M}{r}+\frac{3M\ell_0^2}{r^3}+\mathcal{O}(r^{-5}),
\end{equation}
which shows that the spacetime is asymptotically Schwarzschild, with the
leading quantum correction entering only at order $r^{-3}$. By contrast, close
to the center,
\begin{equation}
 f(r)=1-\frac{2M}{\ell_0^3}r^2+\mathcal{O}(r^4)
 =1-\frac{\Lambda_{\mathrm{eff}}}{3}r^2+\mathcal{O}(r^4),
\end{equation}
where
\begin{equation}
 \Lambda_{\mathrm{eff}}=\frac{6M}{\ell_0^3}.
\end{equation}
Therefore the Schwarzschild singularity is replaced by a de Sitter core, which
is the key manifestation of regularity in this geometry
\cite{Nicolini2019,Konoplya2023}.

The horizons are determined by the roots of $f(r)=0$. If $r_+$ denotes the
outer event horizon, then the horizon condition can be written as
\begin{equation}\label{eq:mass_horizon_relation}
 M=\frac{\left(r_+^2+\ell_0^2\right)^{3/2}}{2r_+^2},
\end{equation}
which is often convenient when one wishes to use the horizon radius rather than
$M$ as the fundamental scale \cite{Konoplya2023}. For fixed positive mass, the
black-hole sector exists only in the range
\begin{equation}\label{eq:parameter_range_mass_units}
 0\leq \frac{\ell_0}{M}\leq \frac{4}{3\sqrt{3}}.
\end{equation}
The lower endpoint reproduces the Schwarzschild geometry, while the upper
endpoint corresponds to the extremal configuration where the inner and outer
horizons merge. In that case,
\begin{equation}
 \ell_0^{\mathrm{ext}}=\frac{4M}{3\sqrt{3}},
 \qquad
 r_{\mathrm{ext}}=\sqrt{2}\,\ell_0^{\mathrm{ext}}
 =\sqrt{\frac{32}{27}}\,M.
\end{equation}
For $\ell_0/M>4/(3\sqrt{3})$, the spacetime remains everywhere regular but no
longer possesses an event horizon. If instead one normalizes all quantities by
$r_+$, then Eq.~\eqref{eq:mass_horizon_relation} implies the equivalent
black-hole range
\begin{equation}
 0\leq \frac{\ell_0}{r_+}\leq \frac{1}{\sqrt{2}}.
\end{equation}

In what follows, we shall be interested exclusively in the black-hole branch,
namely in configurations with $M>0$ and parameter values satisfying
Eq.~\eqref{eq:parameter_range_mass_units}. In this sector, increasing
$\ell_0$ strengthens the deviation from Schwarzschild predominantly in the
near-horizon and interior regions, while leaving the far-zone geometry only
weakly modified. This feature makes the metric especially suitable for the
present analysis of gravitational QNMs and GBFs,
since both observables are controlled by the effective scattering potential in
precisely those regions where the quantum correction is most pronounced.

\section{Axial gravitational perturbations}\label{sec:axial}

{\color{black}
In the Nicolini interpretation, the background is an effective quantum-corrected
geometry. In an effective Einstein description the quantum correction is
encoded in an anisotropic stress-energy tensor, as in the approaches of
Refs.~\cite{Ashtekar:2018lag,Ashtekar:2018cay,Bouhmadi-Lopez:2020oia,KonoplyaStashko2025}.
The geometry of Sec.~\ref{sec:metric} corresponds to an effective
anisotropic fluid with nonvanishing density,
}
\begin{equation}
T^{\mu}{}_{\nu}=\mathrm{diag}\left(-\rho_{\rm eff},p_{r,\rm eff},p_{t,\rm eff},p_{t,\rm eff}\right),
\end{equation}
{\color{black}
The effective source is already part of the original construction of
Ref.~\cite{Nicolini2019}. Its Eq.~(5) explicitly gives the quantum-smeared
density obtained from the regularized Newtonian potential. In Sec.~II the
authors introduce the anisotropic energy-momentum tensor and determine its
pressures from covariant conservation together with $g_{tt}g_{rr}=-1$,
leading to the Bardeen metric in their Eq.~(8). Thus the density and pressures
below agree with the effective source used to encode the quantum corrections
in that construction.

At the perturbative level we adopt an additional, nondissipative effective-fluid
closure: no independent axial perturbation of the unit radial anisotropy
direction $s_\mu$ or additional axial shear stress is introduced, so that
$\delta s_A=0$. Within this closure the angular conservation equation makes
the covariant velocity variable $U$ defined below time independent; hence
$\delta u_A=0$ for the nonzero-frequency modes considered here. Metric-induced
stress perturbations are retained: $\delta T_{aA}=p_{t,\rm eff}h_{aA}$,
where $a\in\{t,r\}$ and $A$ is an angular index. These are the covariant,
or ``down,'' matter conditions of Ref.~\cite{Chakraborty:2024qhm}, whose
axial potential coincides with Eq.~\eqref{eq:axial_potential_fluid} below.
The subsequent quasinormal-mode and grey-body-factor calculations therefore
describe this specified effective perturbation model. The static T-duality
construction fixes the background source but does not uniquely determine its
complete nonspherical response; establishing a unique microscopic
quantum-gravity prediction would require the corresponding linearized
dynamical equations.
}

For a static line element {\eqref{metric}} one has
\begin{align*}
8\pi\rho_{\rm eff}(r)
&=\frac{1-f(r)-r f'(r)}{r^2},
\\
8\pi p_{r,\rm eff}(r)
&=\frac{-1+f(r)+r f'(r)}{r^2}=-8\pi\rho_{\rm eff}(r),
\\
8\pi p_{t,\rm eff}(r)
&=\frac{r f''(r)+2 f'(r)}{2r}.
\end{align*}
Substituting Eq.~\eqref{eq:bardeen_metric_function} into these relations gives
precisely the effective density and pressures used below, namely
\begin{align}
\rho_{\rm eff}(r)
&=\frac{3M\ell_0^2}{4\pi\left(r^2+\ell_0^2\right)^{5/2}},
\\
p_{r,\rm eff}(r)
&=-\rho_{\rm eff}(r),
\\
p_{t,\rm eff}(r)
&=\frac{3M\ell_0^2\left(3r^2-2\ell_0^2\right)}{8\pi\left(r^2+\ell_0^2\right)^{7/2}}.
\end{align}
These are therefore effective source functions associated with the geometry, not
additional matter fields postulated independently of the Nicolini construction.
This point is important for the perturbation problem. As emphasized in the
recent analysis of compact objects supported by anisotropic matter,
different prescriptions for the matter sector can lead to inequivalent
odd-parity spectra, so the treatment of gravitational perturbations must be
stated explicitly rather than inferred from the metric alone
\cite{Chakraborty:2024qhm}. In what follows we therefore formulate the axial
problem in the effective Einstein description with the above nonzero density and
pressures kept throughout the derivation.

{Notice that the tangential pressure may be larger than the energy density, implying that the effective matter content is not physically realistic. While this would indeed be problematic if the black-hole metric were interpreted as an exact solution of the Einstein equations sourced by some supposedly physical matter distribution, this is not the case within the effective approach adopted here. Similar violations of the strong  energy conditions also arise for loop quantum gravity-inspired black holes when an effective energy-momentum tensor is introduced, as demonstrated by Ashtekar \textit{et al.} \cite{Ashtekar:2018cay}. In the next section we will give a full analysis of energy conditions for this configurations.}

We adopt the Regge--Wheeler gauge and expand the odd-parity metric
perturbations in axial vector harmonics \cite{Regge:1957td}. For each
multipole $(\ell,m)$ with $\ell\ge2$, the nonvanishing components are
\begin{align}
 h^{(-)}_{aB}&=\sum_{\ell m}h_a^{\ell m}(t,r)S_B^{\ell m},
\\
 h^{(-)}_{ab}&=0,
\qquad
 h^{(-)}_{AB}=0,
\end{align}
where $a,b\in\{t,r\}$, $A,B\in\{\theta,\phi\}$, and
\begin{equation}
S_A^{\ell m}=\left(-\frac{1}{\sin\theta}\partial_{\phi}Y_{\ell m},\;\sin\theta\,\partial_{\theta}Y_{\ell m}\right)
\end{equation}
are the odd-parity vector spherical harmonics. Since $\rho_{\rm eff}$,
$p_{r,\rm eff}$, and $p_{t,\rm eff}$ are scalars on the two-sphere, they have
no axial perturbations. The only odd-parity matter degree of freedom is the
angular part of the fluid velocity, which we denote by
\begin{equation}
\delta u_A=U^{\ell m}(t,r)S_A^{\ell m}.
\end{equation}
The perturbed conservation law for the effective stress tensor implies that this
axial fluid variable is time independent. Therefore, for the QNM
problem with harmonic dependence $e^{-i\omega t}$ and $\omega\neq0$, it can be
consistently set to zero. In this sense the odd-parity sector reduces to a
single master field, but the background density and radial pressure still enter
nontrivially through the linearized Einstein equations.

To derive the master equation it is useful to introduce the mass function
\begin{equation}
 m(r)=\frac{Mr^3}{\left(r^2+\ell_0^2\right)^{3/2}},
 \qquad
 f(r)=1-\frac{2m(r)}{r},
\end{equation}
so that the background Einstein equations give
\begin{equation}\label{eq:mprime_density}
 m'(r)=4\pi r^2\rho_{\rm eff}(r).
\end{equation}
After substituting the odd-parity ansatz into the $tB$ and $rB$ components of
the linearized Einstein equations, eliminating $h_0$ in favor of $h_1$, and
using the background equations to remove derivatives of the metric, one obtains
for each multipole a Regge--Wheeler-type equation for
\begin{equation}
\Psi(r)=\frac{f(r)}{r}h_1(r)
\end{equation}
in terms of the tortoise coordinate
\begin{equation}
\frac{dr_*}{dr}=\frac{1}{f(r)}.
\end{equation}
The master equation takes the form
\begin{equation}\label{eq:rw_master_density}
\frac{d^2\Psi}{dr_*^2}+\left[\omega^2-V_{\rm ax}(r)\right]\Psi=0,
\end{equation}
with effective potential
\begin{equation}\label{eq:axial_potential_fluid}
\begin{split}
V_{\rm ax}(r)=f(r)\Biggl[\frac{\ell(\ell+1)}{r^2}-\frac{6m(r)}{r^3}
+4\pi\Bigl(\rho_{\rm eff}(r)-p_{r,\rm eff}(r)\Bigr)\Biggr].
\end{split}
\end{equation}
Equation~\eqref{eq:axial_potential_fluid} is the form appropriate to an
anisotropic effective fluid with nonzero density, matching the odd-parity
master potential used for static spherically symmetric anisotropic-fluid
backgrounds in Ref.~\cite{Chakraborty:2024qhm}; see also
Ref.~\cite{Boonserm:2013dua} for the corresponding dirty-black-hole
Regge--Wheeler framework. This is the point at which our derivation differs
conceptually from a purely geometric treatment. {Notice that a complete derivation of the perturbation equations for an anisotropic-fluid source under the most general assumptions, namely without specifying a particular equation of state, can be found in \cite{Chakraborty:2024qhm}. In this framework, two distinct classes of axial perturbations arise, corresponding to different choices of observer: one for which the contravariant components of the fluid four-velocity remain unperturbed, and another for which the covariant components are assumed not to vary.}

For the quantum-corrected Bardeen background the effective equation of state is
special, because $p_{r,\rm eff}=-\rho_{\rm eff}$. Hence
\begin{equation}
V_{\rm ax}(r)=f(r)\left[\frac{\ell(\ell+1)}{r^2}-\frac{6m(r)}{r^3}+8\pi\rho_{\rm eff}(r)\right].
\end{equation}
Substituting Eqs.~\eqref{eq:mprime_density} and the explicit expressions for
$m(r)$ and $\rho_{\rm eff}(r)$, we obtain
\begin{align}
V_{\rm ax}(r)
&=f(r)\left[\frac{\ell(\ell+1)}{r^2}-\frac{6M}{\left(r^2+\ell_0^2\right)^{3/2}}+\frac{6M\ell_0^2}{\left(r^2+\ell_0^2\right)^{5/2}}\right]
\\
&=f(r)\left[\frac{\ell(\ell+1)}{r^2}-\frac{6Mr^2}{\left(r^2+\ell_0^2\right)^{5/2}}\right].
\label{eq:axial_potential_bardeen_final}
\end{align}
Thus, for this particular background, the explicit density term combines with
the mass-function term into the compact potential
\eqref{eq:axial_potential_bardeen_final}. This simplification is a consequence
of the specific relation $p_{r,\rm eff}=-\rho_{\rm eff}$ and should not be
interpreted as a general equivalence between different prescriptions for matter
perturbations.

The resulting potential has the expected limits. At the event horizon,
$f(r_+)=0$, so $V_{\rm ax}(r_+)=0$. In the Schwarzschild limit,
$\ell_0\to0$, Eq.~\eqref{eq:axial_potential_bardeen_final} reduces to the
standard Regge--Wheeler barrier,
\begin{equation}
V_{\rm RW}(r)=\left(1-\frac{2M}{r}\right)\left[\frac{\ell(\ell+1)}{r^2}-\frac{6M}{r^3}\right].
\end{equation}
At large $r$ one has
\begin{equation}
V_{\rm ax}(r)=\frac{\ell(\ell+1)}{r^2}-\frac{2M\left[\ell(\ell+1)+3\right]}{r^3}+\Order{r^{-4}},
\label{eq:axial_potential_asymptotic}
\end{equation}
so the asymptotic solutions of Eq.~\eqref{eq:rw_master_density} are plane waves,
$\Psi\sim e^{\pm i\omega r_*}$, and the usual QNM and scattering
boundary conditions apply.

We emphasize that Eq.~\eqref{eq:axial_potential_bardeen_final} is the axial
potential that follows from the nonvacuum, anisotropic-fluid interpretation of
the Nicolini geometry. In the subsequent sections this is the potential we use
for both the QNM calculation and the GBF analysis.

{\color{black}
The source prescription also distinguishes this calculation from earlier
analyses of gravitational perturbations in the Bardeen family.
Ref.~\cite{Saleh2018} considered a quintessence environment and derived its
perturbation equation using $\delta R_{\mu\nu}=0$.
Refs.~\cite{Dey2019,Zhao2024} instead worked with the nonlinear-electrodynamic
interpretation and a cosmological constant; their axial potentials contain
explicit contributions from the nonlinear-electromagnetic Lagrangian.
The present asymptotically flat problem uses the effective-fluid response
specified here, with metric-induced stress perturbations retained.
In the asymptotically flat, environment-free limit, identifying the background
magnetic parameter with $\ell_0$ identifies the Bardeen lapse functions but
does not, by itself, identify the perturbation operators.
A quantitative comparison requires matching both the matter response and the
boundary conditions. Likewise, the axial/polar comparison of
Ref.~\cite{Zhao2024} addresses a question beyond our axial sector.
}

\begin{figure*}[t]
\centering
\includegraphics[width=\textwidth]{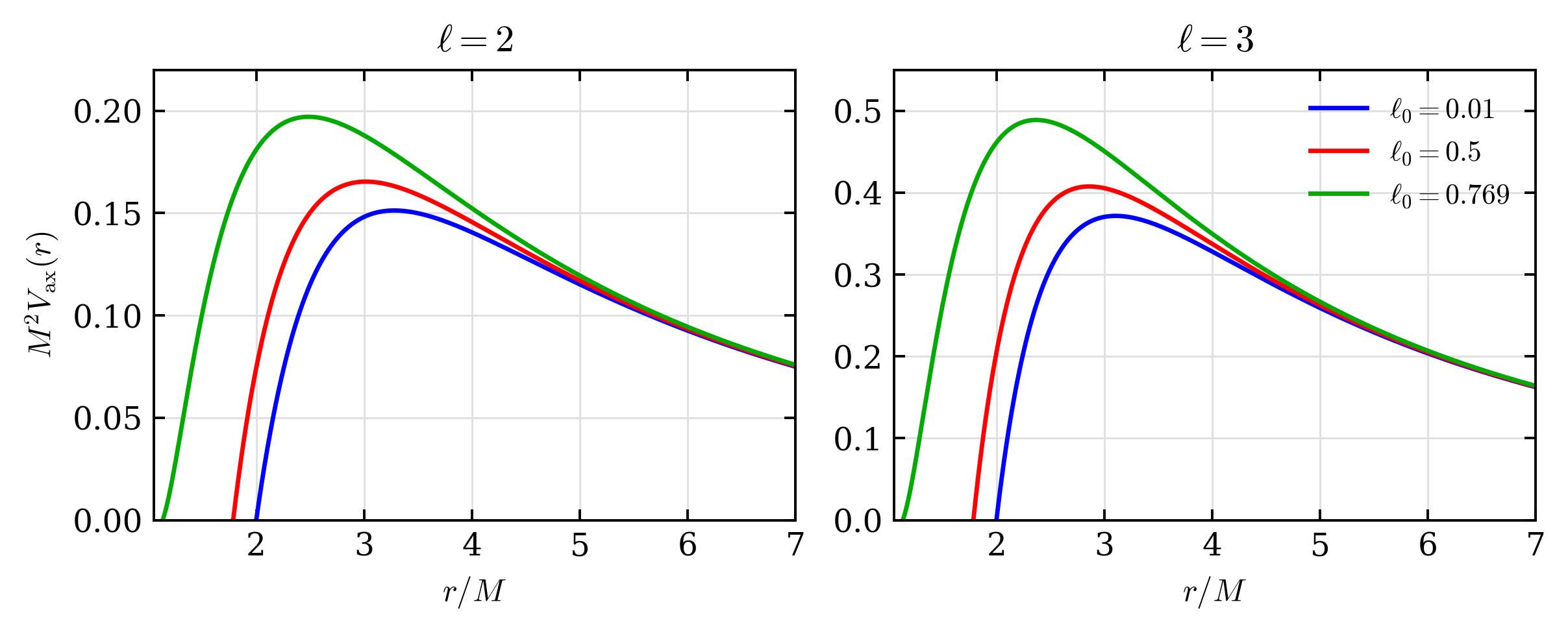}
\caption{Representative axial effective potentials $M^2V_{\rm ax}(r)$ as functions of $r/M$ for the quantum-corrected Bardeen black hole with $M=1$. Panel (a) corresponds to $\ell=2$ and panel (b) to $\ell=3$. In each panel, the blue, red, and green curves denote $\ell_0=0.01$, $0.5$, and $0.769$, respectively.}
\label{fig:axial_potential_profiles}
\end{figure*}

Representative profiles of the potential \eqref{eq:axial_potential_bardeen_final} are shown in Fig.~\ref{fig:axial_potential_profiles} for $M=1$ and for the same values $\ell_0=0.01$, $0.5$, and $0.769$ that will later be used in the scattering analysis. For fixed $\ell$, increasing $\ell_0$ raises the barrier and shifts its maximum toward smaller $r/M$, while increasing $\ell$ produces the expected centrifugal enhancement of the barrier. These trends already indicate that larger $\ell$ and larger $\ell_0$ should postpone the onset of transmission and move the relevant scattering and absorption structures to higher frequencies.

\section{Energy conditions}\label{sec:energy}
{{\color{black}
Let us analyze in more detail what happens with the energy conditions for the obtained effective energy-momentum tensor.
} Since
\begin{equation}
\rho+p_r=0,
\qquad
\rho+p_t=
\frac{15M\ell_0^2r^2}
{8\pi(r^2+\ell_0^2)^{7/2}}
\ge0,
\end{equation}
both the null and weak energy conditions are satisfied everywhere, with the radial null energy condition saturated.}

{The strong energy condition is determined by
\begin{equation}
\rho+p_r+2p_t=2p_t,
\end{equation}
which changes sign at
\begin{equation}
r=\sqrt{\frac23}\,\ell_0.
\end{equation}
Therefore, the strong energy condition is violated only in the central region,
\begin{equation}
r<\sqrt{\frac23}\,\ell_0,
\end{equation}
while it is satisfied outside this radius. Since the event horizon of the black-hole branch always satisfies $r_+\ge\sqrt2\,\ell_0$, the violation of the strong energy condition is confined to the regular core hidden behind the event horizon.}

{Finally, the dominant energy condition requires $\rho\ge|p_r|$ and $\rho\ge|p_t|$. The first inequality is saturated because $p_r=-\rho$, whereas
\begin{equation}
\rho-p_t=
\frac{3M\ell_0^2(4\ell_0^2-r^2)}
{8\pi(r^2+\ell_0^2)^{7/2}},
\end{equation}
showing that the dominant energy condition holds only for
\begin{equation}
r\le2\ell_0
\end{equation}
and is violated at larger distances. Indeed,
\begin{equation}
\frac{p_t}{\rho}
=
\frac{3r^2-2\ell_0^2}{2(r^2+\ell_0^2)}
\longrightarrow
\frac32,
\qquad r\rightarrow\infty,
\end{equation}
so that the tangential pressure asymptotically exceeds the energy density, although both quantities decay as $r^{-5}$. Such a violation of the dominant energy condition is a characteristic feature of the effective matter supporting the geometry and does not indicate the presence of any physical matter source.}

{\color{black}
Its relation to the event horizon follows from
Eq.~\eqref{eq:mass_horizon_relation}:
\[
r_+=2\ell_0\quad\Longleftrightarrow\quad
\frac{\ell_0}{M}=\frac{8}{5\sqrt5}=0.71554175\ldots.
\]
For $0<\ell_0/M\leq8/(5\sqrt5)$ the entire open exterior $r>r_+$
violates the dominant energy condition. Above this threshold, within the
black-hole range, it holds in the shell $r_+\leq r\leq2\ell_0$ and fails
farther out. The Schwarzschild limit $\ell_0=0$ is vacuum and is excluded
from this division. These background energy conditions do not by themselves
determine the linear stability of the specified axial perturbation model.
}

{Once again, since we are dealing with an effective energy-momentum tensor introduced to model gravitational perturbations, the violation of some of the energy conditions does not imply any physical inconsistency. }

\section{WKB method}\label{sec:wkb}

To compute the quasinormal frequencies we employ the standard WKB treatment of
barrier-penetration problems for black-hole perturbations. The first-order
version was introduced by Schutz and Will \cite{SchutzWill1985}, then extended
to third order by Iyer and Will \cite{IyerWill1987}, and later pushed to sixth
order by Konoplya \cite{Konoplya2003WKB}. In its modern implementation, the
method is further improved by Pad\'e resummation of the high-order WKB series,
following the strategy advocated by Matyjasek and Opala
\cite{MatyjasekOpala2017}. This is the approach adopted in the present work.

For the axial master equation \eqref{eq:rw_master_density} it is convenient to
write
\begin{equation}
\frac{d^2\Psi}{dr_*^2}+Q(r_*)\Psi=0,
\qquad
Q(r_*)\equiv \omega^2-V_{\rm ax}(r).
\end{equation}
Quasinormal modes are defined by the boundary conditions of a purely ingoing
wave at the event horizon and a purely outgoing wave at spatial infinity,
\begin{equation}
\Psi\sim
\begin{cases}
 e^{-i\omega r_*}, & r_*\to -\infty,\\
 e^{+i\omega r_*}, & r_*\to +\infty.
\end{cases}
\end{equation}
The WKB construction applies when the effective potential forms a smooth,
single-peaked barrier. For the black-hole branch considered here and for the
low-lying modes studied below, the axial potential satisfies precisely this
condition.

Let $r_{*0}$ denote the position of the maximum of the barrier, so that
$Q_0\equiv Q(r_{*0})$ and
$Q_0^{(p)}\equiv\left.d^pQ/dr_*^p\right|_{r_{*0}}$. Expanding $Q(r_*)$ in a
Taylor series about $r_{*0}$, matching the WKB solutions across the turning
points, and imposing the quasinormal boundary conditions yield the usual WKB
quantization condition
\begin{equation}\label{eq:wkb_quantization}
\frac{iQ_0}{\sqrt{2Q_0^{(2)}}}-\sum_{k=2}^{N}\Lambda_k
= n+\frac12,
\end{equation}
where $n=0,1,2,\ldots$ is the overtone number and the quantities
$\Lambda_k$ are higher-order correction terms built from the derivatives of
$Q$ at the peak. At first order one keeps only the leading term in
Eq.~\eqref{eq:wkb_quantization}, the Iyer--Will extension incorporates the
second- and third-order corrections $\Lambda_2$ and $\Lambda_3$, and the
Konoplya formula includes all corrections up to $\Lambda_6$
\cite{SchutzWill1985,IyerWill1987,Konoplya2003WKB}. In general, an
$N$th-order WKB formula involves peak derivatives up to $Q_0^{(2N)}$; thus the
14th- and 16th-order calculations quoted in our tables require derivatives of
$Q$ up to the 28th and 32nd orders, respectively.

For the present Bardeen background the effective potential is frequency
independent, so the WKB procedure can be implemented directly. We first locate
the peak $r_0$ of $V_{\rm ax}(r)$ outside the event horizon by solving
$dV_{\rm ax}/dr=0$. Since $dr_*/dr=1/f(r)$ and $f(r)>0$ for $r>r_+$, the maxima
in the coordinates $r$ and $r_*$ coincide. The derivatives with respect to the
tortoise coordinate are then obtained from
$d/dr_*=f(r)\,d/dr$ and evaluated at $r=r_0$. Equation
\eqref{eq:wkb_quantization} is subsequently solved numerically for each pair
$(\ell,n)$, and the physically relevant branch is selected by the damping
condition $\im\omega<0$.

A substantial improvement in accuracy is obtained by treating the WKB result as
a formal asymptotic series for $\omega^2$ and resumming it by a Pad\'e
approximant rather than using the truncated polynomial directly
\cite{MatyjasekOpala2017}. Introducing an auxiliary bookkeeping parameter
$\epsilon$, one writes
\begin{equation}\label{eq:wkb_series}
\begin{split}
\omega^2(\epsilon)=&\,V_0-i\mathcal{K}\sqrt{2Q_0^{(2)}}\,\epsilon
-i\sqrt{2Q_0^{(2)}}\sum_{k=2}^{N}\epsilon^k\Lambda_k,\\
&\mathcal{K}=n+\frac12,
\end{split}
\end{equation}
with $V_0\equiv V_{\rm ax}(r_0)$, and only at the end sets $\epsilon=1$.
One then constructs the rational function
\begin{equation}\label{eq:pade_definition}
P_{\tilde n}^{\tilde m}(\epsilon)=
\frac{\sum_{j=0}^{\tilde n}A_j\epsilon^j}
{1+\sum_{j=1}^{\tilde m}B_j\epsilon^j},
\end{equation}
whose Taylor expansion reproduces Eq.~\eqref{eq:wkb_series} through the chosen
order. In the present work we use diagonal Pad\'e approximants. Therefore, the
notation WKB14 $(\tilde m=7)$ in the tables means that the 14th-order WKB
series for $\omega^2$ has been resummed by the diagonal Pad\'e approximant
$P_7^7(\epsilon)$, while WKB16 $(\tilde m=8)$ denotes the 16th-order series
resummed by $P_8^8(\epsilon)$. The agreement between these two high-order,
diagonal Pad\'e estimates provides an internal consistency check on the quoted
quasinormal frequencies.

As in the standard literature, the WKB method is most reliable for smooth,
single-barrier potentials and for low overtones, typically when $n<\ell$ or,
more cautiously, when $n$ is not too close to the top of the available angular
spectrum \cite{Konoplya2003WKB,MatyjasekOpala2017}. Its accuracy improves with
increasing $\ell$ and generally deteriorates for large overtones, for broad or
non-single-peaked barriers, and near situations in which the turning-point
picture becomes marginal.
{\color{black}
Examples of WKB applications include charged, dilatonic and
nonlinear-electromagnetic perturbation problems
\cite{Konoplya:2001ji,Breton:2017hwe,Fernando:2016ftj,Guo:2020caw,Wongjun:2019ydo},
regular geometries and long-lived modes
\cite{Bolokhov:2025egl,Malik:2026lfj,Lutfuoglu:2025ljm,Bolokhov:2025lnt,Pathrikar:2025gzu,Skvortsova:2024eqi},
and modified gravitational backgrounds
\cite{Konoplya:2006ar,Konoplya:2010vz,Momennia:2018hsm,Abdalla:2005hu,Konoplya:2005sy,Sekhmani:2026aun}.
Resummation methods and nonstandard geometries are discussed in
Refs.~\cite{Eniceicu:2019npi,Ishihara:2008re,Konoplya:2018ala,Bolokhov:2025aqy};
related applications concern transmission, radiation and conductivity
\cite{Konoplya:2019ppy,Karmakar:2023cwg,Konoplya:2009hv,Skvortsova2025,Kokkotas:2010zd,Malik:2025erb}.
}
For this reason we use the method below primarily for
the fundamental mode and the first few overtones, precisely the regime in which
it is known to give the most accurate black-hole quasinormal frequencies.

{It is important to note that, strictly speaking, the conventional WKB method converges only asymptotically, and increasing the WKB order alone does not guarantee improved accuracy. However, the application of Pad\'e approximants makes it possible to formulate convergence in an averaged sense. As recently demonstrated in \cite{Konoplya:2026rjh}, the high-order WKB series exhibits Ces\`aro convergence at sufficiently high orders. Although in the present work we neither employed such high WKB orders nor explicitly tested Ces\`aro convergence, the condition $\ell \geq n$ is known to provide very good accuracy, as confirmed by comparisons with the time-domain integration method.}

\section{Time-domain integration method}\label{sec:time}

Although the numerical values quoted below are obtained with the WKB--Pad\'e
approach, it is useful to summarize the standard time-domain evolution scheme
that is commonly used to extract the ringdown signal directly from the master
equation and to monitor the transition from quasinormal ringing to late-time
tails. Introducing the null coordinates
\begin{equation}
 u=t-r_*,
 \qquad
 v=t+r_*,
\end{equation}
the time-dependent version of Eq.~\eqref{eq:rw_master_density} becomes
\begin{equation}
 4\,\frac{\partial^2\Psi}{\partial u\,\partial v}
 +V_{\rm ax}\bigl(r(u,v)\bigr)\Psi=0.
\end{equation}
For black-hole perturbations, a convenient characteristic discretization of
this equation was employed by Gundlach, Price, and Pullin in their analysis of
quasinormal ringing and late-time tails \cite{GundlachPricePullin1994}. If
$S=(u,v)$, $E=(u,v+\Delta)$, $W=(u+\Delta,v)$, and
$N=(u+\Delta,v+\Delta)$ denote the corners of a null grid cell, then the field
update is
\begin{equation}
 \Psi_N=\Psi_W+\Psi_E-\Psi_S
 -\frac{\Delta^2}{8}\,V_{\rm ax}(S)\bigl(\Psi_W+\Psi_E\bigr)
 +\mathcal{O}(\Delta^4).
\end{equation}
One usually specifies a Gaussian pulse on one initial null ray and trivial data
on the other, then records the waveform at fixed $r_*$ as a function of the
time coordinate $t=(u+v)/2$.

The quasinormal part of the signal can then be fitted by a superposition of
damped exponentials. For equally spaced samples
$x_n\equiv\Psi(t_0+n\Delta t)$, the classical Prony procedure
\cite{deProny1795} assumes the representation
\begin{equation}
 x_n=\sum_{j=1}^{p} C_j z_j^{\,n},
 \qquad
 z_j=e^{-i\omega_j\Delta t},
\end{equation}
so that the complex frequencies follow from the roots $z_j$ of the associated
linear-prediction polynomial via
\begin{equation}
 \omega_j=\frac{i}{\Delta t}\ln z_j.
\end{equation}
This provides a direct way to extract the dominant quasinormal frequencies from
the numerically evolved waveform once the initial transient has passed and
before the power-law tail takes over. {\color{black}
Time-domain studies address massive-field tails and long-lived signals
\cite{Konoplya:2023fmh,Bolokhov:2024ixe,Varghese:2011ku,Lutfuoglu:2025hjy,Bolokhov:2026dzn},
stability and dimension-dependent perturbation dynamics
\cite{Skvortsova:2023zca,Konoplya:2013sba,Skvortsova:2023zmj,Abdalla:2012si},
and quantum-motivated or extreme regular black holes
\cite{Skvortsova:2024wly,Malik:2024nhy,Skvortsova:2024atk,Bolokhov:2023bwm}.
Matter environments and modified-gravity ringdown have also been examined
\cite{Bolokhov:2023dxq,Malik:2023bxc,Bolokhov:2026eqf,Dubinsky:2025fwv,Hamil:2025pte}.
Related higher-curvature spectra and scattering problems are discussed in
Refs.~\cite{Arbelaez:2026eaz,Arbelaez:2025gwj,Konoplya:2024hfg,Dubinsky:2024nzo}.
}
For a broad review of time-domain
evolution, late-time tails, and other numerical as well as semianalytical
methods used in black-hole perturbation theory, see
Ref.~\cite{KonoplyaZhidenko2011}. 

\begin{figure*}[t]
\centering
\begin{minipage}{0.49\textwidth}
\centering
\textbf{(a)}\\[-2pt]
\includegraphics[width=\textwidth]{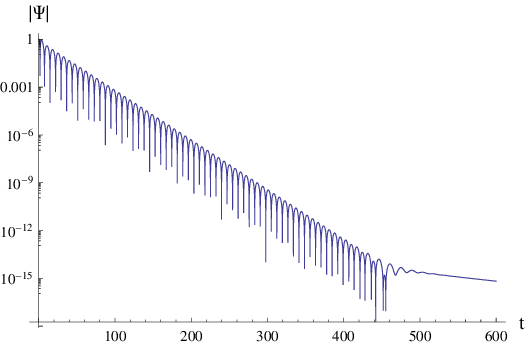}
\end{minipage}\hfill
\begin{minipage}{0.49\textwidth}
\centering
\textbf{(b)}\\[-2pt]
\includegraphics[width=\textwidth]{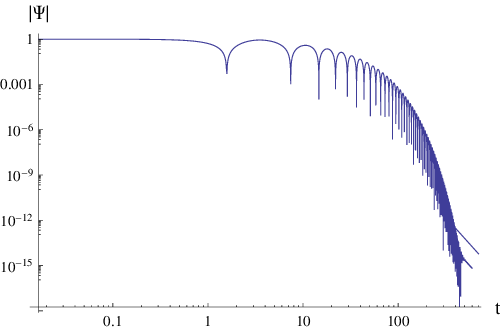}
\end{minipage}
\caption{Time-domain profile of the axial gravitational perturbation for the Bardeen black hole with $\ell=2$, $\ell_0=0.769$, and $M=1$. Panel (a) shows the semilogarithmic profile of $|\Psi|$, displaying the quasinormal ringing and the onset of the late-time tail. Panel (b) shows the same signal on log-log axes, where the asymptotic decay $|\Psi|\propto t^{-7}$ is visible. The fundamental frequency extracted by the Prony method coincides with the WKB--Pad\'e value for the same parameters.}
\label{fig:td_l2}
\end{figure*}

\section{Gravitational quasinormal modes}\label{sec:qnm}

\begin{table}
\centering
\footnotesize
\setlength{\tabcolsep}{3.5pt}
\renewcommand{\arraystretch}{0.95}
\begin{tabular}{c c c c}
\hline
$\ell_0$ & WKB16 ($\tilde{m}=8$) & WKB14 ($\tilde{m}=7$) & Difference \\
\hline
\multicolumn{4}{c}{$\ell=2$, $n=0$} \\
\hline
$0.1$ & $0.374370-0.088885 i$ & $0.374360-0.088869 i$ & $0.0047\%$\\
$0.15$ & $0.375275-0.088706 i$ & $0.375232-0.088740 i$ & $0.0142\%$\\
$0.2$ & $0.376476-0.088543 i$ & $0.376471-0.088551 i$ & $0.00254\%$\\
$0.25$ & $0.378099-0.088296 i$ & $0.378097-0.088298 i$ & $0.0008\%$\\
$0.3$ & $0.380135-0.087969 i$ & $0.380135-0.087969 i$ & $0.0001\%$\\
$0.35$ & $0.382616-0.087551 i$ & $0.382616-0.087551 i$ & $0\%$\\
$0.4$ & $0.385589-0.087021 i$ & $0.385589-0.087021 i$ & $0.00005\%$\\
$0.45$ & $0.389111-0.086351 i$ & $0.389111-0.086351 i$ & $0.\times 10^{\text{-4}}\%$\\
$0.5$ & $0.393261-0.085497 i$ & $0.393261-0.085497 i$ & $0\%$\\
$0.55$ & $0.398144-0.084395 i$ & $0.398144-0.084395 i$ & $0.00002\%$\\
$0.6$ & $0.403901-0.082946 i$ & $0.403902-0.082946 i$ & $0.0001\%$\\
$0.65$ & $0.410728-0.080978 i$ & $0.410728-0.080978 i$ & $0\%$\\
$0.7$ & $0.418895-0.078183 i$ & $0.418895-0.078183 i$ & $0.00006\%$\\
$0.75$ & $0.428728-0.073932 i$ & $0.428728-0.073932 i$ & $0.00002\%$\\
$0.769$ & $0.432944-0.071693 i$ & $0.432944-0.071693 i$ & $0.00003\%$\\
\hline
\multicolumn{4}{c}{$\ell=2$, $n=1$} \\
\hline
$0.1$ & $0.347522-0.273538 i$ & $0.347687-0.273765 i$ & $0.0634\%$\\
$0.15$ & $0.348714-0.273090 i$ & $0.348843-0.273259 i$ & $0.0481\%$\\
$0.2$ & $0.350379-0.272452 i$ & $0.350458-0.272547 i$ & $0.0278\%$\\
$0.25$ & $0.352528-0.271594 i$ & $0.352561-0.271633 i$ & $0.0115\%$\\
$0.3$ & $0.355189-0.270474 i$ & $0.355204-0.270495 i$ & $0.0057\%$\\
$0.35$ & $0.358400-0.269057 i$ & $0.358415-0.269079 i$ & $0.0058\%$\\
$0.4$ & $0.362243-0.267257 i$ & $0.362250-0.267262 i$ & $0.00185\%$\\
$0.45$ & $0.366756-0.264963 i$ & $0.366758-0.264960 i$ & $0.0008\%$\\
$0.5$ & $0.371997-0.262057 i$ & $0.372001-0.262054 i$ & $0.0011\%$\\
$0.55$ & $0.378069-0.258315 i$ & $0.378070-0.258315 i$ & $0.0001\%$\\
$0.6$ & $0.385043-0.253413 i$ & $0.385047-0.253419 i$ & $0.0015\%$\\
$0.65$ & $0.392954-0.246806 i$ & $0.392954-0.246807 i$ & $0.\times 10^{\text{-4}}\%$\\
$0.7$ & $0.401596-0.237556 i$ & $0.401596-0.237550 i$ & $0.0013\%$\\
$0.75$ & $0.409771-0.224243 i$ & $0.409772-0.224240 i$ & $0.00060\%$\\
$0.769$ & $0.412108-0.217950 i$ & $0.412110-0.217947 i$ & $0.00089\%$\\
\hline
\multicolumn{4}{c}{$\ell=2$, $n=2$} \\
\hline
$0.1$ & $0.301922-0.477390 i$ & $0.303284-0.475856 i$ & $0.363\%$\\
$0.15$ & $0.303648-0.476330 i$ & $0.304523-0.475374 i$ & $0.229\%$\\
$0.2$ & $0.306056-0.474876 i$ & $0.306558-0.474375 i$ & $0.126\%$\\
$0.25$ & $0.309149-0.473015 i$ & $0.309458-0.472755 i$ & $0.0715\%$\\
$0.3$ & $0.312956-0.470779 i$ & $0.313473-0.470515 i$ & $0.103\%$\\
$0.35$ & $0.317624-0.467931 i$ & $0.317739-0.468124 i$ & $0.0397\%$\\
$0.4$ & $0.323011-0.464232 i$ & $0.323026-0.464370 i$ & $0.0245\%$\\
$0.45$ & $0.329187-0.459525 i$ & $0.329265-0.459916 i$ & $0.0707\%$\\
$0.5$ & $0.336158-0.453558 i$ & $0.329541-0.459080 i$ & $1.53\%$\\
$0.55$ & $0.344078-0.445962 i$ & $0.344008-0.445701 i$ & $0.0480\%$\\
$0.6$ & $0.352825-0.436132 i$ & $0.352822-0.436119 i$ & $0.00236\%$\\
$0.65$ & $0.361328-0.423254 i$ & $0.361950-0.423071 i$ & $0.117\%$\\
$0.7$ & $0.369751-0.405120 i$ & $0.369858-0.405181 i$ & $0.0225\%$\\
$0.75$ & $0.372329-0.382540 i$ & $0.372319-0.382510 i$ & $0.00603\%$\\
$0.769$ & $0.371221-0.373784 i$ & $0.371219-0.373930 i$ & $0.0277\%$\\
\hline
\end{tabular}
\caption{Gravitational quasinormal frequencies for the Bardeen black hole with $\ell=2$ and $M=1$, shown as functions of $\ell_0$ for the fundamental mode ($n=0$) and the first two overtones ($n=1,2$). The second and third columns list the 16th- and 14th-order WKB--Pad\'e values, while the last column gives the percentage difference between the two approximations.}
\label{tab:qnm_l2}
\end{table}

\begin{table}
\centering
\footnotesize
\setlength{\tabcolsep}{3.5pt}
\renewcommand{\arraystretch}{0.95}
\begin{tabular}{c c c c}
\hline
$\ell_0$ & WKB16 ($\tilde{m}=8$) & WKB14 ($\tilde{m}=7$) & Difference \\
\hline
\multicolumn{4}{c}{$\ell=3$, $n=0$} \\
\hline
$0.1$ & $0.600497-0.092602 i$ & $0.600497-0.092602 i$ & $0\%$\\
$0.15$ & $0.601828-0.092472 i$ & $0.601828-0.092472 i$ & $0\%$\\
$0.2$ & $0.603718-0.092284 i$ & $0.603718-0.092284 i$ & $0\%$\\
$0.25$ & $0.606198-0.092030 i$ & $0.606198-0.092030 i$ & $0\%$\\
$0.3$ & $0.609306-0.091698 i$ & $0.609306-0.091698 i$ & $0\%$\\
$0.35$ & $0.613096-0.091275 i$ & $0.613096-0.091275 i$ & $0\%$\\
$0.4$ & $0.617638-0.090739 i$ & $0.617638-0.090739 i$ & $0\%$\\
$0.45$ & $0.623026-0.090061 i$ & $0.623026-0.090061 i$ & $0\%$\\
$0.5$ & $0.629381-0.089197 i$ & $0.629381-0.089197 i$ & $0\%$\\
$0.55$ & $0.636871-0.088083 i$ & $0.636871-0.088083 i$ & $0\%$\\
$0.6$ & $0.645725-0.086618 i$ & $0.645725-0.086618 i$ & $0\%$\\
$0.65$ & $0.656272-0.084631 i$ & $0.656272-0.084631 i$ & $0\%$\\
$0.7$ & $0.669002-0.081808 i$ & $0.669002-0.081808 i$ & $0\%$\\
$0.75$ & $0.684670-0.077495 i$ & $0.684670-0.077495 i$ & $0\%$\\
$0.769$ & $0.691612-0.075187 i$ & $0.691612-0.075187 i$ & $0\%$\\
\hline
\multicolumn{4}{c}{$\ell=3$, $n=1$} \\
\hline
$0.1$ & $0.583823-0.280973 i$ & $0.583823-0.280973 i$ & $0\%$\\
$0.15$ & $0.585311-0.280555 i$ & $0.585311-0.280555 i$ & $0\%$\\
$0.2$ & $0.587424-0.279950 i$ & $0.587424-0.279950 i$ & $0\%$\\
$0.25$ & $0.590191-0.279132 i$ & $0.590191-0.279132 i$ & $0\%$\\
$0.3$ & $0.593654-0.278069 i$ & $0.593654-0.278069 i$ & $0\%$\\
$0.35$ & $0.597868-0.276713 i$ & $0.597868-0.276713 i$ & $0\%$\\
$0.4$ & $0.602904-0.274998 i$ & $0.602904-0.274998 i$ & $0\%$\\
$0.45$ & $0.608853-0.272832 i$ & $0.608853-0.272832 i$ & $0\%$\\
$0.5$ & $0.615833-0.270079 i$ & $0.615833-0.270079 i$ & $0\%$\\
$0.55$ & $0.623994-0.266537 i$ & $0.623994-0.266537 i$ & $0.00002\%$\\
$0.6$ & $0.633524-0.261890 i$ & $0.633524-0.261890 i$ & $0\%$\\
$0.65$ & $0.644646-0.255614 i$ & $0.644646-0.255614 i$ & $0\%$\\
$0.7$ & $0.657548-0.246774 i$ & $0.657548-0.246774 i$ & $0\%$\\
$0.75$ & $0.672003-0.233598 i$ & $0.672003-0.233598 i$ & $0\%$\\
$0.769$ & $0.677555-0.226871 i$ & $0.677555-0.226871 i$ & $0\%$\\
\hline
\end{tabular}
\caption{Gravitational quasinormal frequencies for the Bardeen black hole with $\ell=3$ and $M=1$, shown as functions of $\ell_0$ for the fundamental mode ($n=0$) and the first overtone ($n=1$). The second and third columns list the 16th- and 14th-order WKB--Pad\'e values, while the last column gives the percentage difference between the two approximations.}
\label{tab:qnm_l3_low}
\end{table}

\begin{table}
\centering
\footnotesize
\setlength{\tabcolsep}{3.5pt}
\renewcommand{\arraystretch}{0.95}
\begin{tabular}{c c c c}
\hline
$\ell_0$ & WKB16 ($\tilde{m}=8$) & WKB14 ($\tilde{m}=7$) & Difference \\
\hline
\multicolumn{4}{c}{$\ell=3$, $n=2$} \\
\hline
$0.1$ & $0.553098-0.478475 i$ & $0.553099-0.478473 i$ & $0.00032\%$\\
$0.15$ & $0.554881-0.477681 i$ & $0.554882-0.477680 i$ & $0.00014\%$\\
$0.2$ & $0.557407-0.476531 i$ & $0.557407-0.476531 i$ & $0\%$\\
$0.25$ & $0.560709-0.474983 i$ & $0.560709-0.474983 i$ & $0.00006\%$\\
$0.3$ & $0.564830-0.472976 i$ & $0.564829-0.472975 i$ & $0.00009\%$\\
$0.35$ & $0.569824-0.470424 i$ & $0.569824-0.470423 i$ & $0.00010\%$\\
$0.4$ & $0.575762-0.467209 i$ & $0.575762-0.467208 i$ & $0.00009\%$\\
$0.45$ & $0.582728-0.463162 i$ & $0.582728-0.463162 i$ & $0.00007\%$\\
$0.5$ & $0.590821-0.458042 i$ & $0.590821-0.458041 i$ & $0.00009\%$\\
$0.55$ & $0.600148-0.451486 i$ & $0.600149-0.451486 i$ & $0.00017\%$\\
$0.6$ & $0.610797-0.442935 i$ & $0.610797-0.442935 i$ & $0\%$\\
$0.65$ & $0.622736-0.431480 i$ & $0.622737-0.431480 i$ & $0.00008\%$\\
$0.7$ & $0.635472-0.415600 i$ & $0.635473-0.415600 i$ & $0.00005\%$\\
$0.75$ & $0.646798-0.393129 i$ & $0.646798-0.393128 i$ & $0.00004\%$\\
$0.769$ & $0.649593-0.382667 i$ & $0.649593-0.382666 i$ & $0.00003\%$\\
\hline
\multicolumn{4}{c}{$\ell=3$, $n=3$} \\
\hline
$0.1$ & $0.513672-0.689321 i$ & $0.513679-0.689307 i$ & $0.00181\%$\\
$0.15$ & $0.515827-0.688013 i$ & $0.515832-0.688006 i$ & $0.00088\%$\\
$0.2$ & $0.518878-0.686124 i$ & $0.518880-0.686122 i$ & $0.00038\%$\\
$0.25$ & $0.522853-0.683588 i$ & $0.522852-0.683589 i$ & $0.00016\%$\\
$0.3$ & $0.527797-0.680313 i$ & $0.527792-0.680314 i$ & $0.00049\%$\\
$0.35$ & $0.533758-0.676166 i$ & $0.533755-0.676167 i$ & $0.00039\%$\\
$0.4$ & $0.540798-0.670966 i$ & $0.540795-0.670966 i$ & $0.00039\%$\\
$0.45$ & $0.548982-0.664455 i$ & $0.548975-0.664454 i$ & $0.00083\%$\\
$0.5$ & $0.558367-0.656262 i$ & $0.558359-0.656253 i$ & $0.00142\%$\\
$0.55$ & $0.568971-0.645840 i$ & $0.568979-0.645836 i$ & $0.00108\%$\\
$0.6$ & $0.580690-0.632341 i$ & $0.580685-0.632342 i$ & $0.00056\%$\\
$0.65$ & $0.593023-0.614459 i$ & $0.593024-0.614453 i$ & $0.00069\%$\\
$0.7$ & $0.604255-0.590244 i$ & $0.604256-0.590242 i$ & $0.00027\%$\\
$0.75$ & $0.609422-0.558724 i$ & $0.609355-0.558799 i$ & $0.0121\%$\\
$0.769$ & $0.608341-0.545838 i$ & $0.608340-0.545837 i$ & $0.00002\%$\\
\hline
\end{tabular}
\caption{Gravitational quasinormal frequencies for the Bardeen black hole with $\ell=3$ and $M=1$, shown as functions of $\ell_0$ for the second and third overtones ($n=2,3$). The second and third columns list the 16th- and 14th-order WKB--Pad\'e values, while the last column gives the percentage difference between the two approximations.}
\label{tab:qnm_l3_high}
\end{table}

\begin{table}
\centering
\footnotesize
\setlength{\tabcolsep}{3.5pt}
\renewcommand{\arraystretch}{0.95}
\begin{tabular}{c c c c}
\hline
$\ell$ & WKB16 ($\tilde{m}=8$) & WKB14 ($\tilde{m}=7$) & Difference \\
\hline
$2$ & $0.432944-0.071693 i$ & $0.432944-0.071693 i$ & $0.00003\%$\\
$3$ & $0.691612-0.075187 i$ & $0.691612-0.075187 i$ & $0\%$\\
$4$ & $0.931790-0.076402 i$ & $0.931790-0.076402 i$ & $0\%$\\
$5$ & $1.164521-0.076967 i$ & $1.164521-0.076967 i$ & $0\%$\\
$6$ & $1.393465-0.077276 i$ & $1.393465-0.077276 i$ & $0\%$\\
$7$ & $1.620206-0.077465 i$ & $1.620206-0.077465 i$ & $0\%$\\
$8$ & $1.845549-0.077589 i$ & $1.845549-0.077589 i$ & $0\%$\\
$9$ & $2.069948-0.077675 i$ & $2.069948-0.077675 i$ & $0\%$\\
$10$ & $2.293678-0.077737 i$ & $2.293678-0.077737 i$ & $0\%$\\
$11$ & $2.516916-0.077783 i$ & $2.516916-0.077783 i$ & $0\%$\\
$12$ & $2.739783-0.077819 i$ & $2.739783-0.077819 i$ & $0\%$\\
$13$ & $2.962363-0.077846 i$ & $2.962363-0.077846 i$ & $0\%$\\
$14$ & $3.184714-0.077868 i$ & $3.184714-0.077868 i$ & $0\%$\\
$15$ & $3.406883-0.077886 i$ & $3.406883-0.077886 i$ & $0\%$\\
$16$ & $3.628902-0.077901 i$ & $3.628902-0.077901 i$ & $0\%$\\
$17$ & $3.850796-0.077913 i$ & $3.850796-0.077913 i$ & $0\%$\\
$18$ & $4.072588-0.077924 i$ & $4.072588-0.077924 i$ & $0\%$\\
$19$ & $4.294292-0.077932 i$ & $4.294292-0.077932 i$ & $0\%$\\
$20$ & $4.515921-0.077940 i$ & $4.515921-0.077940 i$ & $0\%$\\
\hline
\end{tabular}
\caption{Fundamental gravitational quasinormal frequencies ($n=0$) for the Bardeen black hole with $M=1$ and fixed $\ell_0=0.769$, shown as functions of the multipole number $\ell$. The second and third columns list the 16th- and 14th-order WKB--Pad\'e values, while the last column gives the percentage difference between the two approximations.}
\label{tab:qnm_lscan}
\end{table}

The quasinormal spectra collected in Tables~\ref{tab:qnm_l2}--\ref{tab:qnm_l3_high}
show a clear and systematic dependence on the quantum-correction parameter
$\ell_0$. For every mode listed, increasing $\ell_0$ shifts the spectrum toward
larger oscillation frequencies and weaker damping, i.e., $\re\omega$ increases
while $|\im\omega|$ decreases. For the fundamental $\ell=2$ mode, for example,
$\omega$ changes from $0.374370-0.088885\,i$ at $\ell_0=0.1$ to
$0.432944-0.071693\,i$ at $\ell_0=0.769$, whereas for the fundamental
$\ell=3$ mode it changes from $0.600497-0.092602\,i$ to
$0.691612-0.075187\,i$ over the same interval. The same qualitative trend is
seen for the first few overtones: at fixed $\ell$, higher overtones remain more
damped than the fundamental mode, but increasing $\ell_0$ still moves them
upward in $\re\omega$ and closer to the real axis. Thus the quantum-corrected
Bardeen black hole rings both faster and longer than its near-Schwarzschild
counterpart.

The multipole dependence is summarized in Table~\ref{tab:qnm_lscan}, where the
fundamental mode is displayed for fixed $\ell_0=0.769$. As $\ell$ increases
from $2$ to $20$, the real part of the frequency grows from $0.432944$ to
$4.515921$, while the damping rate varies only mildly, from $0.071693$ to
$0.077940$. Therefore the real part grows almost linearly with $\ell$, whereas
$|\im\omega|$ rapidly approaches an approximately constant value. This is the
expected eikonal behavior: higher multipoles oscillate much more rapidly, but
their damping time changes comparatively little.

This trend can be made quantitative in the eikonal regime. For large
$L\equiv\ell+1/2$, the axial potential
\eqref{eq:axial_potential_bardeen_final} takes the form
\begin{equation}
 V_{\rm ax}(r)=L^2\,\frac{f(r)}{r^2}+\Order{L^0}.
\end{equation}
Hence the peak of the leading barrier is determined by 
\begin{equation}\label{eq:eikonal_peak_condition}
\begin{aligned}
 \left.\frac{d}{dr}\left(\frac{f(r)}{r^2}\right)\right|_{r=r_{\rm ph}}=0
 \\
 &\Longleftrightarrow\quad 2f(r_{\rm ph})=r_{\rm ph}f'(r_{\rm ph}),
\end{aligned}
\end{equation}
that is, by the outer unstable null orbit. For the Bardeen metric
\eqref{eq:bardeen_metric_function}, this condition becomes
\begin{equation}\label{eq:bardeen_photon_orbit}
 3Mr_{\rm ph}^4=\left(r_{\rm ph}^2+\ell_0^2\right)^{5/2}.
\end{equation}
Keeping only the leading term in the WKB formula
\eqref{eq:wkb_quantization} and expanding in inverse powers of $L$, one finds
the gravitational counterpart of the test-field eikonal formula discussed in
Ref.~\cite{Konoplya2023},
\begin{equation}\label{eq:bardeen_eikonal_qnm}
\begin{aligned}
 \omega_{\ell n}&=L\,\Omega_c-i\left(n+\frac12\right)\lambda_c+\Order{L^{-1}},
 \\
 L&\equiv\ell+\frac12,
\end{aligned}
\end{equation}
where
\begin{align}
 \Omega_c^2&=\frac{f(r_{\rm ph})}{r_{\rm ph}^2}
 =\frac{r_{\rm ph}^2-2\ell_0^2}{3r_{\rm ph}^4}, \\
 \lambda_c^2&=\frac{(r_{\rm ph}^2-4\ell_0^2)(r_{\rm ph}^2-2\ell_0^2)}
 {3r_{\rm ph}^4(r_{\rm ph}^2+\ell_0^2)} \\
 &=\frac{f(r_{\rm ph})}{2r_{\rm ph}^2}
 \left[2f(r_{\rm ph})-r_{\rm ph}^2f''(r_{\rm ph})\right].
\end{align}
For the fixed value $\ell_0=0.769$ used in Table~\ref{tab:qnm_lscan},
the outer solution of Eq.~\eqref{eq:bardeen_photon_orbit} is
$r_{\rm ph}\simeq2.30360\,M$, which gives
$\Omega_c M\simeq0.220941$ and $\lambda_c M\simeq0.156021$. {\color{black}
Here $\ell=2,3,\ldots$ is the integer angular multipole, while $\ell_0$
is the dimensionful metric parameter of Eq.~\eqref{metric} and
Eq.~\eqref{eq:bardeen_metric_function}. We write the two mode indices
with a comma when the overtone is specified: the second index in
$\omega_{\ell,0}$ is $n=0$, the fundamental overtone, and is unrelated
to the subscript in $\ell_0$. Thus the fundamental photon-sphere mode has
\begin{equation}\label{eq:eikonal_n0_numeric}
 \omega_{\ell,0}M \simeq 0.220941\left(\ell+\frac12\right)-0.078011\,i.
\end{equation}
}
Comparison with Table~\ref{tab:qnm_lscan} shows that the eikonal formula
converges monotonically from above. In the real part, the deviation decreases
from $4.35\%$ at $\ell=5$ to $1.14\%$ at $\ell=10$ and $0.30\%$ at
$\ell=20$; for $|\im\omega|$, the corresponding deviations are
$1.36\%$, $0.35\%$, and $0.09\%$. Thus the table confirms the analytic
eikonal law very well once $\ell\gtrsim10$: the damping rate is already in
the sub-percent regime for moderate multipoles, while the oscillation frequency
approaches the same level slightly more slowly.

{\color{black}
The rate of approach to the eikonal limit is quantified for several
geometries in Table~\ref{tab:eikonal_convergence}. For the $n=0$
photon-sphere branch we define the componentwise relative discrepancies
\begin{align*}
\delta_R&=100\frac{|L\Omega_c-\operatorname{Re}\omega_{\ell,0}|}
{|\operatorname{Re}\omega_{\ell,0}|},\\
\delta_I&=100\frac{|\lambda_c/2-|\operatorname{Im}\omega_{\ell,0}||}
{|\operatorname{Im}\omega_{\ell,0}|}.
\end{align*}
The tabulated frequencies use WKB--Pad\'e $P_8^8$ at order $16$.
Order-$14$ $P_7^7$ values provide an internal check. Their maximum
complex-frequency spread in this table is below $10^{-9}\%$; this internal
agreement is distinct from the measured error of the leading eikonal formula.
For all four geometries, increasing $\ell$ from $5$ to $20$ reduces the
real-part discrepancy from about $4.4$--$4.6\%$ to $0.30$--$0.31\%$,
and the damping-rate discrepancy from about $1.4\%$ to below $0.10\%$.

This approximately $L^{-2}$ relative convergence has a simple origin. The
potential can be written exactly as $V_{\rm ax}=L^2F+G$, where
$F=f/r^2$ and
$G=f[-1/(4r^2)-6Mr^2/(r^2+\ell_0^2)^{5/2}]$.
For the nondegenerate outer maximum,
\[
r_0=r_{\rm ph}-\frac{G'(r_{\rm ph})}{L^2F''(r_{\rm ph})}
+\mathcal{O}(L^{-4}).
\]
The peak shift is therefore suppressed by $L^{-2}$, and the fixed-overtone expansion has
$\operatorname{Re}\omega=L\Omega_c+\mathcal{O}(L^{-1})$ and
$|\operatorname{Im}\omega|=\lambda_c/2+\mathcal{O}(L^{-2})$
for $n=0$, giving relative corrections of order $L^{-2}$ to both components.

The same behavior persists in the near-extremal example
$\ell_0/M=0.7697$, for which $r_+/M=1.1038738$,
while $r_{\rm ph}/M=2.3014844$ and $M\lambda_c=0.1558238$.
The outer photon-sphere barrier remains separated from the horizon and
nondegenerate. At the exact metric extremal value,
$\ell_0/M=4/(3\sqrt3)$, its limiting values are
$M\Omega_c=0.2210405$ and $M\lambda_c=0.1557954$.
Our numerical test concerns the fixed-$n$ photon-sphere branch selected by
the barrier WKB calculation; it does not establish uniform accuracy in the
extremal limit or identify any distinct near-horizon mode branches.
}
\begin{table*}[t]
\centering\color{black}\small
\setlength{\tabcolsep}{5pt}
\begin{tabular}{rrcccc}
\hline\hline
$\ell_0/M$ & $\ell$ & $M\operatorname{Re}\omega_{\ell,0}$ & $-M\operatorname{Im}\omega_{\ell,0}$ & $\delta_R\;(\%)$ & $\delta_I\;(\%)$\\
\hline
$0$ & $5$ & $1.01229531$ & $0.09487052$ & $4.56193$ & $1.42777$ \\
$0$ & $10$ & $1.99678779$ & $0.09586387$ & $1.19883$ & $0.37676$ \\
$0$ & $20$ & $3.93300139$ & $0.09613107$ & $0.31084$ & $0.09776$ \\
\hline
$0.5$ & $5$ & $1.06142023$ & $0.09125030$ & $4.46972$ & $1.37466$ \\
$0.5$ & $10$ & $2.09234253$ & $0.09217225$ & $1.17463$ & $0.36065$ \\
$0.5$ & $20$ & $4.12048425$ & $0.09241832$ & $0.30457$ & $0.09344$ \\
\hline
$0.769$ & $5$ & $1.16452137$ & $0.07696674$ & $4.34996$ & $1.35628$ \\
$0.769$ & $10$ & $2.29367753$ & $0.07773703$ & $1.14257$ & $0.35194$ \\
$0.769$ & $20$ & $4.51592091$ & $0.07793975$ & $0.29622$ & $0.09092$ \\
\hline
$0.7697$ & $5$ & $1.16497690$ & $0.07686945$ & $4.35009$ & $1.35616$ \\
$0.7697$ & $10$ & $2.29457705$ & $0.07763871$ & $1.14259$ & $0.35190$ \\
$0.7697$ & $20$ & $4.51769279$ & $0.07784116$ & $0.29623$ & $0.09091$ \\
\hline
\hline\end{tabular}
\caption{Eikonal convergence for the $n=0$ photon-sphere branch of the axial gravitational potential at fixed $M=1$. Frequencies are calculated with sixteenth-order WKB--Pad\'e $P_8^8$; the eikonal estimate is $(\ell+1/2)\Omega_c-i\lambda_c/2$. The componentwise relative discrepancies $\delta_R$ and $\delta_I$ are defined in the text and evaluated before rounding. The last geometry is close to the extremal value $\ell_0/M=4/(3\sqrt3)=0.76980036\ldots$.}\label{tab:eikonal_convergence}
\end{table*}

The internal accuracy of the semianalytic calculation can be assessed from the
comparison between the 16th- and 14th-order WKB--Pad\'e approximants. For the
fundamental mode the agreement is excellent: the discrepancy never exceeds
$0.0142\%$ for $\ell=2$ and is zero at the displayed precision for $\ell=3$.
For the first overtone the maximal mismatch is still only $0.0634\%$ for
$\ell=2$ and $2\times10^{-5}\%$ for $\ell=3$. The higher overtones behave as
expected from the usual WKB domain of validity: the agreement remains very good
for the $\ell=3$ spectrum, while the largest deviation in the whole data set is
$1.53\%$ for the $\ell=2$, $n=2$, $\ell_0=0.5$ mode, where the condition
$n<\ell$ is only marginally satisfied. Independent support comes from the
time-domain profile shown in Fig.~\ref{fig:td_l2}: the Prony extraction of the
fundamental mode for $\ell=2$ and $\ell_0=0.769$ reproduces the same frequency
$\omega=0.432944-0.071693\,i$ as the WKB--Pad\'e calculation.

\begin{figure}[t]
\centering
\includegraphics[width=\columnwidth]{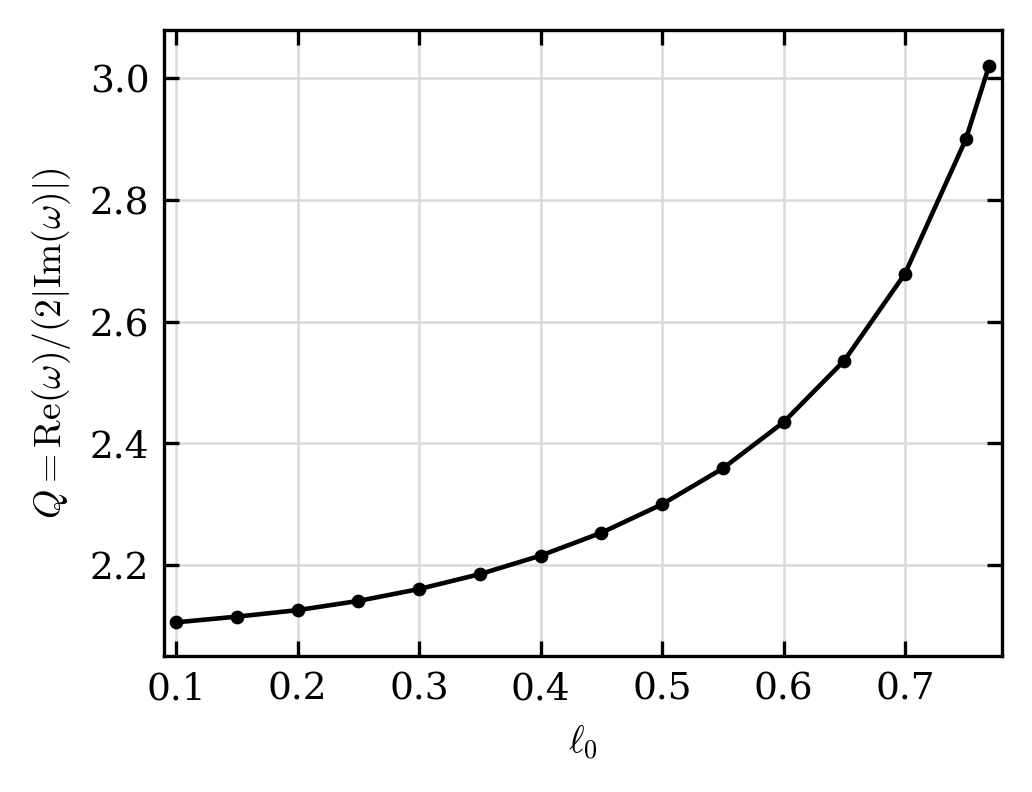}
\caption{Quality factor $Q=\re{\omega}/(2|\im{\omega}|)$ for the fundamental
axial gravitational mode with $\ell=2$, computed from the WKB16 data in
Table~\ref{tab:qnm_l2}. The monotonic increase with $\ell_0$ shows that the
quantum-corrected Bardeen black hole becomes a progressively better oscillator
as one approaches the extremal regime.}
\label{fig:qfactor_l2}
\end{figure}

A convenient measure of how good an oscillator the black hole is is the quality
factor,
\begin{equation}\label{eq:q_factor_definition}
 Q_{\ell n}\equiv \frac{\re\omega_{\ell n}}{2|\im\omega_{\ell n}|}.
\end{equation}
Larger $Q_{\ell n}$ means that the system performs more oscillations before the
signal is substantially damped. For the fundamental $\ell=2$ mode,
Fig.~\ref{fig:qfactor_l2} shows that $Q$ grows monotonically with the
quantum-correction parameter, from $Q\simeq2.106$ at $\ell_0=0.1$ to
$Q\simeq3.019$ at $\ell_0=0.769$. Since the classical Schwarzschild black hole
is recovered in the limit $\ell_0\to0$, this monotonic increase implies that
within this family the quantum-corrected black holes are better oscillators
than the classical one, and the near-extremal configurations are the best
oscillators of all.

The time-domain evolution confirms the same picture. After the initial burst,
the signal first enters the ringdown regime,
\begin{equation}
 \Psi(t,r_*)\approx A\,e^{-i\omega_0 t},
\end{equation}
where $\omega_0$ is the fundamental quasinormal frequency, and only at later
times does it cross over to a power-law tail. Because the far-zone axial
potential behaves as in Eq.~\eqref{eq:axial_potential_asymptotic}, the usual
Price law for asymptotically flat black holes with compact-support initial data
applies at fixed radius,
\begin{equation}\label{eq:price_law_fixed_r}
 \Psi_{\ell}(t,r)\propto t^{-(2\ell+3)},
 \qquad t\to\infty,
 \qquad r={\rm const}.
\end{equation}
Thus, for the present gravitational mode with $\ell=2$, one expects
$\Psi\propto t^{-7}$, exactly as seen in the log-log panel of
Fig.~\ref{fig:td_l2}. In this sense the late-time signal follows the standard
Price law \cite{Price1972,GundlachPricePullin1994,KonoplyaZhidenko2011}. The
important point is that the quantum-correction parameter changes the quasinormal
frequencies appreciably, but it does not change the tail exponent because the
large-$r$ structure of the effective potential remains of the Price-law type.

\section{Grey-body factors}\label{sec:gbf}

{\color{black}
Related investigations have combined estimates of quasinormal frequencies,
grey-body transmission and horizon thermodynamics for static loop-quantum-gravity
black holes \cite{Dong2025} and Simpson--Visser black-bounce spacetimes
\cite{MartinezGuerrero2026}. These examples illustrate the value of examining
several wave and horizon observables together. Here we focus on the axial
gravitational channel of the Bardeen geometry with the effective-source
prescription stated above.
}

Grey-body factors describe the frequency-dependent transmission of waves through
the curvature-induced potential barrier surrounding the black hole. For the
axial gravitational perturbations considered here, one studies the same master
equation \eqref{eq:rw_master_density}, but now with real frequency
$\Omega>0$ and with scattering boundary conditions rather than QNM
boundary conditions. Choosing a unit-amplitude wave incident from spatial
infinity, the asymptotic behavior is
\begin{equation}\label{eq:gbf_boundary_conditions}
\Psi\sim
\begin{cases}
 T_{\ell}(\Omega)e^{-i\Omega r_*}, & r_*\to -\infty,\\
 e^{-i\Omega r_*}+R_{\ell}(\Omega)e^{+i\Omega r_*}, & r_*\to +\infty,
\end{cases}
\end{equation}
where $R_{\ell}(\Omega)$ and $T_{\ell}(\Omega)$ are the reflection and
transmission amplitudes. Since the background is static and the effective
potential is real, the Wronskian is conserved and one has the standard flux
relation
\begin{equation}\label{eq:gbf_flux_conservation}
 |R_{\ell}(\Omega)|^2+|T_{\ell}(\Omega)|^2=1.
\end{equation}
The GBF is therefore defined as
\begin{equation}\label{eq:gbf_definition}
 \Gamma_{\ell}(\Omega)\equiv |T_{\ell}(\Omega)|^2,
\end{equation}
which measures the fraction of a given partial wave that penetrates the
potential barrier. In the scattering picture it is the transmission
probability from infinity to the horizon, while in the Hawking-radiation
picture the same quantity determines how much of the near-horizon flux reaches
an asymptotic observer \cite{Page:1976df,Boonserm:2013dua,Konoplya2023}. Thus,
low frequencies are strongly suppressed by the barrier, whereas when
$\Omega^2\gg V_{\rm ax}^{\rm max}$ the barrier becomes nearly transparent and
$\Gamma_{\ell}(\Omega)\to1$.

\begin{figure*}[t]
\centering
\begin{minipage}[t]{0.49\textwidth}
\centering
\includegraphics[width=\linewidth]{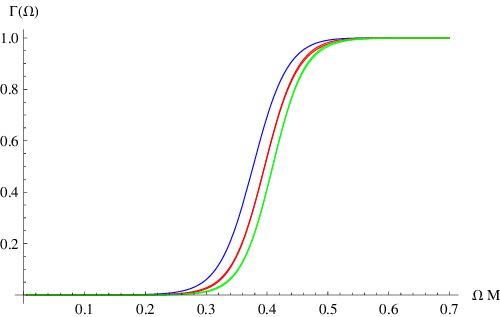}
\par\vspace{-0.8ex}
{\small \textbf{(a)}}
\end{minipage}\hfill
\begin{minipage}[t]{0.49\textwidth}
\centering
\includegraphics[width=\linewidth]{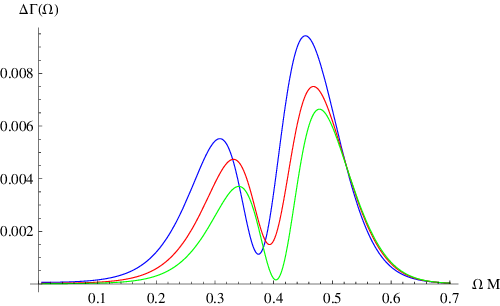}
\par\vspace{-0.8ex}
{\small \textbf{(b)}}
\end{minipage}
\caption{Comparison of the GBFs for axial gravitational perturbations of the quantum-corrected Bardeen black hole with $\ell=2$ and $M=1$. Panel (a) shows the transmission probabilities obtained from the sixth-order WKB method and from the QNM-correspondence formulas for $\ell_0=0.01$ (blue), $\ell_0=0.5$ (red), and $\ell_0=0.769$ (green). Panel (b) shows the absolute difference between the two methods for the same parameter values.}
\label{fig:gbf_compare_l2}
\end{figure*}

\begin{figure*}[t]
\centering
\begin{minipage}[t]{0.49\textwidth}
\centering
\includegraphics[width=\linewidth]{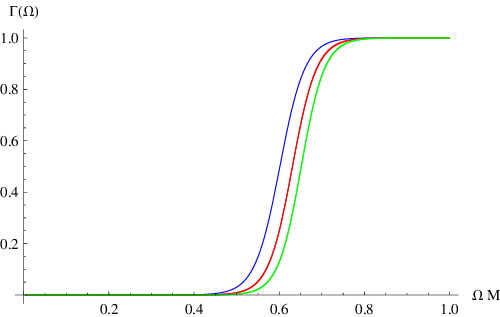}
\par\vspace{-0.8ex}
{\small \textbf{(a)}}
\end{minipage}\hfill
\begin{minipage}[t]{0.49\textwidth}
\centering
\includegraphics[width=\linewidth]{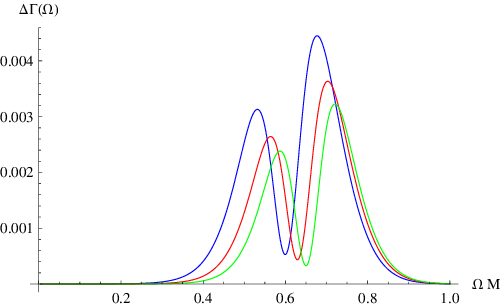}
\par\vspace{-0.8ex}
{\small \textbf{(b)}}
\end{minipage}
\caption{Comparison of the GBFs for axial gravitational perturbations of the quantum-corrected Bardeen black hole with $\ell=3$ and $M=1$. Panel (a) shows the transmission probabilities obtained from the sixth-order WKB method and from the QNM-correspondence formulas for $\ell_0=0.01$ (blue), $\ell_0=0.5$ (red), and $\ell_0=0.769$ (green). Panel (b) shows the absolute difference between the two methods for the same parameter values.}
\label{fig:gbf_compare_l3}
\end{figure*}

\begin{figure*}[t]
\centering
\begin{minipage}[t]{0.49\textwidth}
\centering
\includegraphics[width=\linewidth]{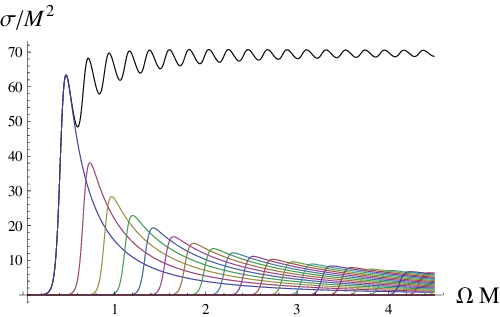}
\par\vspace{-0.8ex}
{\small \textbf{(a)}}
\end{minipage}\hfill
\begin{minipage}[t]{0.49\textwidth}
\centering
\includegraphics[width=\linewidth]{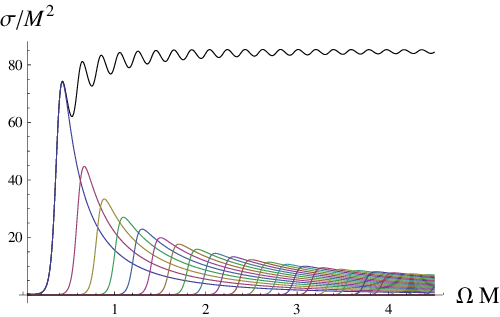}
\par\vspace{-0.8ex}
{\small \textbf{(b)}}
\end{minipage}
\caption{Partial and total ACSs for axial gravitational perturbations of the quantum-corrected Bardeen black hole with $M=1$. Panel (a) corresponds to $\ell_0=0.01$, while panel (b) corresponds to $\ell_0=0.769$. In each panel, the black curve represents the total absorption cross-section and the colored curves represent the partial ACSs of the individual multipoles.}
\label{fig:acs_compare}
\end{figure*}

Because the effective potential \eqref{eq:axial_potential_bardeen_final} is a
smooth, single-peaked barrier, the WKB formalism used for the QNM
problem can be applied to the scattering problem as well. At the lowest WKB
order, the transmission probability is approximated by the familiar
barrier-penetration formula
\begin{equation}\label{eq:gbf_wkb_first_order}
\Gamma_{\ell}(\Omega)\approx
\left[
1+\exp\!\left(
\frac{2\pi\left(V_0-\Omega^2\right)}{\sqrt{-2V_2}}
\right)
\right]^{-1},
\end{equation}
where $V_0\equiv V_{\rm ax}(r_0)$ is the value of the potential at its peak and
$V_2\equiv \left.d^2V_{\rm ax}/dr_*^2\right|_{r_{*0}}<0$. Higher-order
corrections are incorporated by the same local expansion around the peak that
leads to Eq.~\eqref{eq:wkb_quantization}. Following the WKB treatment of the
scattering problem, one introduces a continuous quantity $\mathcal{K}(\Omega)$
and writes the transmission probability as
\begin{equation}\label{eq:gbf_wkb_general}
 \Gamma_{\ell}(\Omega)=\frac{1}{1+e^{2\pi i\mathcal{K}(\Omega)}}.
\end{equation}
Here $\mathcal{K}(\Omega)$ is built from the derivatives of the potential at
its maximum exactly in the same way as in the WKB analysis of quasinormal
modes \cite{IyerWill1987,KonoplyaZhidenko2024Corr}. Unlike in our QNM
calculation, we do not use Pad\'e resummation here. Instead, to make the
GBF--QNM correspondence explicit, we use the asymptotic formulas derived in
Refs.~\cite{KonoplyaZhidenko2024Corr,KonoplyaZhidenko2025Rot} and specialize
them to the present static notation.

The basic point is that the same WKB quantity $\mathcal{K}$ governs both the
scattering problem and the resonance problem. In scattering,
$\mathcal{K}(\Omega)$ varies continuously with the real frequency and enters
Eq.~\eqref{eq:gbf_wkb_general}; in the QNM problem, the WKB
condition requires $\mathcal{K}=n+\tfrac12$, with $n=0,1,2,\ldots$ the
overtone number. Thus the fundamental mode $\omega_0$ and the first overtone
$\omega_1$ encode the same local information about the peak of the potential
barrier that determines the GBF. For asymptotically flat black
holes and large multipole number $\ell$, the correspondence can be written in a
particularly compact form. At leading eikonal order one has
\begin{equation}\label{eq:gbf_qnm_correspondence_eikonal}
 i\mathcal{K}(\Omega)=
 \frac{\Omega^2-\re{\omega_0}^2}{4\re{\omega_0}\im{\omega_0}}
 +\mathcal{O}(\ell^{-1}),
\end{equation}
which shows that, once inserted into Eq.~\eqref{eq:gbf_wkb_general}, the
GBF is determined by the real and imaginary parts of the
fundamental quasinormal frequency alone. In particular, the transition region
between almost complete reflection and almost complete transmission is centered
near $\Omega^2\simeq \re{\omega_0}^2$, while its width is controlled by the
damping rate $|\im{\omega_0}|$.

A more accurate relation, including the second-order correction beyond the
eikonal limit, reads
\begin{widetext}
\begin{equation}\label{eq:gbf_qnm_correspondence_second_order}
\begin{aligned}
 i\mathcal{K}(\Omega)=&\,
 \frac{\Omega^2-\re{\omega_0}^2}{4\re{\omega_0}\im{\omega_0}}
 \left(1+\frac{(\re{\omega_0}-\re{\omega_1})^2}{32\im{\omega_0}^2}
 -\frac{3\im{\omega_0}-\im{\omega_1}}{24\im{\omega_0}}\right)
 -\frac{\re{\omega_0}-\re{\omega_1}}{16\im{\omega_0}}
 \\
 &
 -\frac{(\Omega^2-\re{\omega_0}^2)^2}{16\re{\omega_0}^3\im{\omega_0}}
 \left(1+\frac{\re{\omega_0}(\re{\omega_0}-\re{\omega_1})}{4\im{\omega_0}^2}\right)
 \\
 &
 +\frac{(\Omega^2-\re{\omega_0}^2)^3}{32\re{\omega_0}^5\im{\omega_0}}
 \Bigg(1+\frac{\re{\omega_0}(\re{\omega_0}-\re{\omega_1})}{4\im{\omega_0}^2}
 \\
 &\qquad\qquad
 +\re{\omega_0}^2\left[
 \frac{(\re{\omega_0}-\re{\omega_1})^2}{16\im{\omega_0}^4}
 -\frac{3\im{\omega_0}-\im{\omega_1}}{12\im{\omega_0}}
 \right]\Bigg)+\mathcal{O}(\ell^{-3}).
\end{aligned}
\end{equation}
\end{widetext}
Equation~\eqref{eq:gbf_qnm_correspondence_second_order} makes the
correspondence more explicit: beyond the leading barrier-top approximation, the
shape of the transmission curve depends not only on the fundamental mode but
also on the separation between the fundamental mode and the first overtone.
Since the present Bardeen spacetime is static and has no superradiant sector,
these non-superradiant correspondence formulas apply directly to each axial
gravitational multipole considered here.

The relation between QNMs and GBFs has been investigated extensively in a number of recent works \cite{Malik:2025qnr,Lutfuoglu:2025mqa,Lutfuoglu:2026uzy,Bolokhov:2024otn,Lutfuoglu:2025kqp,Dubinsky:2024vbn,Lutfuoglu:2026gey,Malik:2024wvs,Lutfuoglu:2025ldc,Lutfuoglu:2025eik}, where it was found to provide a reasonably accurate description even for relatively low multipole numbers $\ell$. The validity of this correspondence hinges on the same conditions that underlie the applicability of the WKB approximation, in particular the existence of a single, well-defined potential barrier. When this condition is violated—such as in the case of a double-well structure of the effective potential \cite{Konoplya:2025hgp}, or when higher-curvature corrections significantly alter the centrifugal term and lead to instabilities of the background \cite{Konoplya:2017ymp,Konoplya:2017lhs,Dotti:2004sh,Takahashi:2010gz,Konoplya:2017zwo}—the correspondence is expected to break down. {In our case, as can be seen from representative figures \ref{fig:axial_potential_profiles}, the effective potential has a single maximum.}

For the representative axial mode $\ell=2$ with $M=1$,
Fig.~\ref{fig:gbf_compare_l2} compares the GBFs obtained from the
sixth-order WKB method with those reconstructed from the QNM
correspondence for $\ell_0=0.01$, $0.5$, and $0.769$. The same figure also
shows the absolute difference between the two approaches.

For the representative axial mode $\ell=3$ with $M=1$,
Fig.~\ref{fig:gbf_compare_l3} shows the corresponding comparison between the
sixth-order WKB GBFs and the QNM-correspondence reconstruction
for $\ell_0=0.01$, $0.5$, and $0.769$. As before, we also display the absolute
difference between the two methods.

Figure~\ref{fig:axial_potential_profiles} provides a direct interpretation of Figs.~\ref{fig:gbf_compare_l2} and \ref{fig:gbf_compare_l3}. Because both increasing $\ell$ and increasing $\ell_0$ raise the characteristic barrier height, the condition $\Omega^2\sim V_{\rm ax}^{\rm max}$ is reached only at larger frequencies. Consequently, the transmission curves shift to the right as $\ell_0$ increases, and the $\ell=3$ channels turn on later than the $\ell=2$ channels. The same barrier ordering explains the stronger suppression of the low-frequency GBFs in the more strongly quantum-corrected cases.

The error plots admit the same interpretation. The absolute differences between the sixth-order WKB result and the QNM-correspondence reconstruction are concentrated in the intermediate-frequency window where the barrier changes from mostly reflective to mostly transparent, while they become negligible in the limits $\Gamma_\ell\approx0$ and $\Gamma_\ell\approx1$, where both methods are forced to agree. Moreover, the smaller discrepancies for $\ell=3$ are consistent with the more eikonal character of the higher-multipole barrier, for which the local barrier-top information encoded in $\omega_0$ and $\omega_1$ reconstructs the transmission curve more accurately.

{The other characteristic, which is constructed from grey-body factors at various $\ell$, \textit{the absorption cross section},  quantifies the effective area presented by a black hole for capturing incident radiation. For a given frequency, it is obtained by summing the partial absorption probabilities over all angular multipoles,
\begin{equation}
\sigma(\omega)={\color{black}\sum_{\ell=2}^{\infty}}\frac{\pi(2\ell+1)}{\omega^2}\Gamma_{\ell}(\omega),
\end{equation}
where \(\Gamma_{\ell}(\omega)\) is the grey-body factor for the multipole number \(\ell\). At high frequencies, the absorption cross section approaches the geometric-optics limit, determined by the radius of the photon sphere.} {\color{black}
For Fig.~\ref{fig:acs_compare} we evaluate the partial and multipole-summed
absorption cross-sections of the axial model as
\begin{equation}\label{eq:acs_partial_sum}
\begin{aligned}
 \sigma_\ell^{\rm ax}(\Omega)&=\frac{\pi(2\ell+1)}{\Omega^2}\Gamma_\ell(\Omega),\\
 \sigma_L^{\rm ax}(\Omega)&=\sum_{\ell=2}^{L}\sigma_\ell^{\rm ax}(\Omega).
\end{aligned}
\end{equation}

\begin{table*}[t]
\centering\color{black}
\small\setlength{\tabcolsep}{4pt}
\renewcommand{\arraystretch}{1.12}
\begin{tabular}{rrrrrrr}
\hline\hline
$\ell_0/M$ & $\Omega M$ & $\sigma_{30}^{\rm ax}/M^2$ & $\sigma_{50}^{\rm ax}/M^2$ &
$\sigma_{80}^{\rm ax}/M^2$ & $\sigma_{100}^{\rm ax}/M^2$ &
$\sigma_{\rm geo}/M^2$\\
\hline
$0.01$ & $4.5$ & $84.361992$ & $84.361992$ & $84.361992$ & $84.361992$ & $84.820174$\\
$0.01$ & $8$ & $46.976628$ & $84.980912$ & $84.980912$ & $84.980912$ & $84.820174$\\
$0.01$ & $12$ & $20.878501$ & $56.657751$ & $84.613543$ & $84.613543$ & $84.820174$\\
$0.769$ & $4.5$ & $63.610689$ & $63.610689$ & $63.610689$ & $63.610689$ & $64.357000$\\
$0.769$ & $8$ & $46.976628$ & $63.742631$ & $63.742631$ & $63.742631$ & $64.357000$\\
$0.769$ & $12$ & $20.878501$ & $56.657751$ & $63.974503$ & $63.974503$ & $64.357000$\\
\hline\hline
\end{tabular}
\caption{Direct-integration convergence check of the axial absorption sum at
fixed $M=1$. The first frequency is the upper end of Fig.~\ref{fig:acs_compare};
the two larger frequencies test the approach to the geometric-optics regime.
Each numerical entry is evaluated directly at the stated frequency. The
last column is the photon-capture cross-section of the corresponding metric.
Six decimal places are shown; agreement at that precision tests angular-sum
truncation separately from the radial-integration accuracy.}
\label{tab:acs_convergence}
\end{table*}
Here ``total'' denotes the sum over the axial multipoles studied in this
work. Both panels use $M=1$, with $\ell_0=0.01$ and $0.769$, respectively.
For these absorption curves, the transmission probabilities are calculated
by direct integration of Eq.~\eqref{eq:rw_master_density}, imposing an
ingoing horizon solution and matching to incoming and outgoing waves in the
far region. This calculation supplies an independent check of the absorption
sum, while the comparisons in Figs.~\ref{fig:gbf_compare_l2}
and~\ref{fig:gbf_compare_l3} retain the WKB and QNM-correspondence methods.

The direct calculation uses a sixth-order regular horizon expansion and
a sixteenth-order far-field expansion for the plotted curves, with relative
integration tolerance $2\times10^{-11}$. The higher-multipole cutoff checks use a
thirty-second-order far-field expansion. The outer matching radius and
far-field order are varied to check convergence.
Selected transmission probabilities were repeated with tighter tolerances,
an enlarged matching radius, and a reduced horizon offset; the changes were
below $10^{-8}$ in relative value. Flux conservation was checked independently.
The plotted curves use $224$ frequency samples: $\Omega M=0.05$ and $0.06,0.08,\ldots,4.50$. The interpolation of individual
partial cross-sections is checked against additional directly integrated
points to an absolute tolerance of $10^{-6}M^2$ per partial wave.

We retain $\ell=2,\ldots,30$ over the displayed interval
$0.05\leq\Omega M\leq4.5$. The cutoff test in
Table~\ref{tab:acs_convergence} compares $L=30,50,80,100$ using direct
transmission probabilities, without interpolation. At $\Omega M=4.5$, all four sums agree to the six decimal places shown. The explicit contribution of $\ell=31,\ldots,100$ at that endpoint is below $10^{-8}$ of the total for both geometries.
The omitted high multipoles are below the accuracy of the plotted curves;
increasing the cutoff therefore preserves their oscillatory structure.
The oscillations reflect the successive onset of transmission in individual
partial waves, rather than a termination of the angular-momentum sum.

The geometric-optics limit is fixed by the outer unstable photon orbit:
\begin{equation}\label{eq:acs_capture_limit}
\begin{aligned}
 \sigma_{\rm geo}&=\pi b_c^2,\qquad
 b_c^2=\frac{r_{\rm ph}^2}{f(r_{\rm ph})},\\
 2f(r_{\rm ph})&=r_{\rm ph}f'(r_{\rm ph}).
\end{aligned}
\end{equation}
For $\ell_0/M=0.01$ and $0.769$, respectively, this gives
$r_{\rm ph}/M=2.9999167$ and $2.3035976$, and
$\sigma_{\rm geo}/M^2=84.820174$ and $64.357000$.
The converged sums oscillate about these limits with decreasing amplitude
as the frequency increases. A fixed cutoff cannot be used at arbitrarily
large frequency: once all retained channels are transparent,
$\sigma_L^{\rm ax}\simeq\pi[(L+1)^2-4]/\Omega^2$.
Thus the higher-frequency checks require a larger $L$, as illustrated by
the $L=50$ deficit at $\Omega M=12$ in Table~\ref{tab:acs_convergence}.

At fixed mass the near-extremal capture area is therefore $24.1\%$ smaller.
The near-extremal total cross-section is lower throughout the displayed
frequency interval. At low frequency the higher axial
barrier reduces the transmission of a fixed multipole; at high frequency the
smaller critical impact parameter reduces the number of efficiently absorbed
partial waves, approximately $\ell+1/2\lesssim\Omega b_c$.
These two effects explain the lower total cross-section and the shifted
partial-wave peaks in Fig.~\ref{fig:acs_compare}.
}

\section{Conclusions}\label{sec:conclusions}

In this work we have studied axial gravitational perturbations of the asymptotically flat Bardeen spacetime in its string-T-duality-inspired quantum-corrected interpretation. Starting from the nonvacuum anisotropic-fluid form of the background, we derived the Regge--Wheeler-type master equation and obtained the compact effective potential \eqref{eq:axial_potential_bardeen_final}. The resulting potential is a smooth single barrier that vanishes at the event horizon, reduces to the standard Regge--Wheeler potential in the Schwarzschild limit, and becomes both higher and more inward-shifted as the quantum-correction parameter $\ell_0$ increases. This barrier deformation provides the common physical mechanism behind the QNM, GBF, and ACS results obtained in the paper.

Using high-order WKB--Pad\'e techniques, supplemented by time-domain integration and Prony extraction, we found that increasing $\ell_0$ systematically increases the oscillation frequencies and decreases the damping rates of the axial gravitational modes. In other words, the quantum-corrected Bardeen black hole rings faster and for a longer time than its near-Schwarzschild counterpart, and the quality factor of the fundamental mode grows monotonically toward the near-extremal regime. At fixed $\ell_0$, the real part of the fundamental frequency grows almost linearly with the multipole number, whereas the imaginary part approaches an approximately constant eikonal value. The agreement between the 16th- and 14th-order WKB--Pad\'e approximants is excellent for the dominant modes, and the time-domain profile confirms the semianalytic result for the representative $\ell=2$ configuration. At late times, the signal still obeys the standard Price-law decay, showing that the near-horizon quantum correction modifies the ringing spectrum much more strongly than the asymptotic tail exponent.

The scattering observables reveal a closely related pattern. Because the effective barrier becomes higher for larger $\ell$ and larger $\ell_0$, the onset of transmission is shifted to higher frequencies and the low-frequency GBFs are more strongly suppressed. The QNM-correspondence formulas reproduce the sixth-order WKB transmission curves rather well, with the best agreement in the regimes of almost complete reflection or transmission and, as expected, even better performance for the more eikonal $\ell=3$ case. The partial and total ACSs exhibit the same barrier-governed structure: successive partial-wave peaks appear as higher multipoles begin to penetrate the barrier, while the total cross-section inherits the resulting oscillatory pattern. Overall, the results indicate that the quantum-correction parameter leaves a coherent imprint on both ringdown and scattering observables, making the Bardeen geometry a useful phenomenological proxy for exploring how short-distance regularization may influence gravitational spectroscopy and black-hole absorption.

\begin{acknowledgments}
B. C. L. is grateful to the Excellence project FoS UHK 2205/2025-2026 for the financial support.
\end{acknowledgments}

\section*{Conflict of Interest}
The authors declare no conflict of interest.

\section*{Data Availability}
{\color{black}
All numerical data is represented in the current manuscript. Additional numerical data can be provided by authors upon reasonable request.  
}

\bibliography{referencesNEW}

\end{document}